\documentclass[a4paper, onecolumn, accepted=2021-07-15]{quantumarticle}
\pdfoutput=1

\usepackage{hyperref}
\usepackage[square,numbers]{natbib}
\usepackage{braket}
\usepackage{amsmath}
\usepackage{amsfonts}
\usepackage{amssymb}
\usepackage{bm}
\usepackage[T1]{fontenc}

\newcommand{\mi}{\mathrm{i}}

\begin{document}

\title{Spacetime Quantum Reference Frames and superpositions of proper times}

\author{Flaminia Giacomini}%
 \email{fgiacomini@perimeterinstitute.ca}
\affiliation{%
Perimeter Institute for Theoretical Physics, 31 Caroline St. N, Waterloo, Ontario, N2L 2Y5, Canada
}%

\begin{abstract}
	In general relativity, the description of spacetime relies on idealised rods and clocks, which identify a reference frame. In any concrete scenario, reference frames are associated to physical systems, which are ultimately quantum in nature. A relativistic description of the laws of physics hence needs to take into account such quantum reference frames (QRFs), through which spacetime can be given an operational meaning. 

Here, we introduce the notion of a spacetime quantum reference frame, associated to a quantum particle in spacetime. Such formulation has the advantage of treating space and time on equal footing, and of allowing us to describe the dynamical evolution of a set of quantum systems from the perspective of another quantum system, where the parameter in which the rest of the physical systems evolves coincides with the proper time of the particle taken as the QRF. Crucially, the proper times in two different QRFs are not related by a standard transformation, but they might be in a quantum superposition one with respect to the other.	

Concretely, we consider a system of $N$ relativistic quantum particles in a weak gravitational field, and introduce a timeless formulation in which the global state of the $N$ particles appears ``frozen'', but the dynamical evolution is recovered in terms of relational quantities. The position and momentum Hilbert space of the particles is used to fix the QRF via a transformation to the local frame of the particle such that the metric is locally inertial at the origin of the QRF. The internal Hilbert space corresponds to the clock space, which keeps the proper time in the local frame of the particle. Thanks to this fully relational construction we show how the remaining particles evolve dynamically in the relational variables from the perspective of the QRF. The construction proposed here includes the Page-Wootters mechanism for non interacting clocks when the external degrees of freedom are neglected. Finally, we find that a quantum superposition of gravitational redshifts and a quantum superposition of special-relativistic time dilations can be observed in the QRF.
\end{abstract}

\maketitle

\section{Introduction}

In both quantum theory and general relativity, spacetime is treated as an abstract entity. In quantum theory, spacetime is considered as a fixed, non-dynamical background providing an arena for physical phenomena. In general relativity, spacetime is a dynamical quantity, which is influenced by a change in the configuration of massive bodies. In both cases, spacetime coordinates acquire meaning via rods and clocks, which identify a reference frame. However, these rods and clocks, and hence reference frames, are typically treated as idealised classical systems. In this sense, the description of spacetime is not fully operational. 

Reference frames are very useful to specify the point of view from which observations are carried out. Although measurements are usually made in a specific reference frame, all physical laws are formulated in a way that is independent of the reference frame chosen, thanks to the principle of general covariance. This means that there is no preferred reference frame.

The rods and clocks which specify the reference frames are physical systems, and ultimately quantum in nature. Hence, an operational formulation of the laws of physics requires being able to describe physical phenomena from the point of view of such quantum systems, which can be in a superposition or entangled with other physical systems. This is the main intuition behind the notion of Quantum Reference Frames (QRFs).

QRFs have been studied in quantum information, quantum foundations, and quantum gravity since 1967. In the quantum information literature~\cite{aharonov1, aharonov2, aharonov3, brs_review, bartlett_communication, spekkens_resource, kitaev_superselection, palmer_changing, bartlett_degradation, smith_quantumrf, poulin_dynamics, poulin_deformed, skotiniotis_frameness, poulin_reformulation, busch_relational_1, busch_relational_2, busch_relational_3, jacques, angelo_1, angelo_2, angelo_3}, they have been used to overcome superselection rules, applied to quantum communication scenarios, and related to invariance properties and symmetries of quantum systems. In the quantum gravity literature, it has been conjectured that QRFs are needed to formulate a quantum theory of gravity \cite{dewitt1967quantum, rovelli_quantum}, especially when taking a relational perspective on spacetime~\cite{rovelli_relational}. In particular, a relational approach to QRFs is common to both the quantum gravity and quantum information approaches. Recently, Ref.~\cite{QRF} introduced a formalism to describe physics from the point of view of a specific QRF and transformations to change the description between different QRFs. Subsequent work developed this approach further in different contexts. Refs.~\cite{perspective1, perspective2, yang2020switching} adopt a symmetry principle to describe a Galilean system in a fully relational way, and Refs.~\cite{giacomini2019relativistic, streiter2020relativistic} extend the formalism to the special relativistic regime, applying it to concrete problems in Relativistic Quantum Information. Other works~\cite{de2020quantum, krumm2020, ballesteros2020group} have focussed on the group properties of QRFs both in discrete and in continuous-variable systems. A time version of QRFs, also related to the Page-Wootters mechanism~\cite{Page:1983uc, giovannetti2015quantum}, has been introduced in Ref.~\cite{castro2020quantum} for interacting clocks. Recently, a generalisation of the formalism has been applied to to quantum systems living on a superposition of gravitational fields, and showed that the Einstein Equivalence Principle can be extended to QRFs~\cite{giacomini2020einstein}. 

The work on QRFs is related to research on quantum clocks, which have been studied  with a relational approach, both in the interacting and non-interacting case, in Refs.~\cite{hoehn2018switch, hohn2019switching, smith2019quantizing, hoehn2019trinity, hoehn2020equivalence, smith2020quantum}. In particular, Ref.~\cite{smith2020quantum} studied relativistic clocks by extending the Page-Wootters approach to a relativistic particle with internal and external degrees of freedom. A conceptually similar approach to QRFs, but with some important differences, is the quantum coordinate systems~\cite{Hardy:2018kbp, zych2018relativity}.
 
All these works have studied QRFs either in space or in time, and separately considered the role of internal and external variables (such as, respectively, position or momentum and internal quantities acting as clocks) in identifying the QRF. So far, no unifying approach to QRFs in spacetime has been formulated. A spacetime formulation of QRFs is a crucial step towards a fully relational and covariant description of physics from the point of view of a quantum system, and could also provide a natural setting for a quantum approach to time. However, achieving such a formulation faces some important challenges, such as describing space and time in quantum theory on an equal footing. This task requires the incorporation of a time operator into the description of QRFs in space, and is conceptually related to the problem of time in quantum gravity \cite{isham1993canonical, Rovelli:2004tv, kuchavr2011time}. The methodology and the scope of the present work are different to those employed in the canonical approaches to the problem of time.

In this work, we overcome the difficulty of incorporating a time operator into QRFs and develop a model for spacetime QRFs. We introduce a timeless and fully relational formulation, which methodologically adopts techniques from different approaches to QRFs~\cite{QRF, perspective1, perspective2, castro2020quantum, giacomini2020einstein}, unifying them by adopting aspects of Covariant Quantum Mechanics~\cite{rovelli1990quantum, reisenberger2002spacetime, hellmann2007multiple} and the Page-Wootters mechanism~\cite{Page:1983uc, giovannetti2015quantum}. In particular, we consider a set of $N$ relativistic quantum particles in a weak gravitational field, each of which has a quantum state living in the tensor product Hibert space of position/momentum and some internal ``clock'' Hilbert space (see Refs.~\cite{zych2019gravitational, smith2020quantum} for a similar analysis in a different context). In this timeless formulation, the global state of the $N$ particles is ``frozen'', but the dynamical evolution is recovered in terms of the relational variables to one of the particles, which is chosen as the QRF.  In order to obtain this relational description on the reduced set of $N-1$ particles, we build a transformation to some relational variables (e.g., relative positions) of the particle chosen as the QRF, and then fix the origin of the QRF at the location of this particle. The evolution of the remaining $N-1$ particles is parametrised in terms of the proper time of the particle serving as the QRF. Concretely, the proper time of each particle is encoded in its internal variables, which serve as quantum clocks. Hence, both the external and the internal variables play an important role in buiding the spacetime QRF: the former allow us to transform to the local frame of a quantum particle, whose spacetime position can be in a superposition or entangled with other systems, and the latter allow us to describe the dynamical evolution of the remaining particles in terms of the proper time in each QRF. Notice that, although proper time is a standard evolution parameter in each QRF, the proper times of two different quantum particles are not in a classical relation one with respect to the other, but can be in a quantum superposition. This joint use of the external variables to fix the QRF and the internal variables as clocks is one of the novel aspects of this work.

In addition, we show that we can always find a transformation to the QRF of the particle such that the metric is locally inertial at the origin of the QRF. This is an instance of a Quantum Locally Inertial Frame (QLIF), as introduced in Ref.~\cite{giacomini2020einstein}. We find the relational dynamics of the remaining particles from the point of view of the chosen QRF, which coincides, in the appropriate limit, with the dynamical evolution predicted with different methods (for instance, see Refs.~\cite{zych_interferometric, pikovski_TimeDilation,zych, pikovski}). The expression of this relational dynamics has the same functional form in any chosen QRF, and is thus symmetric under QRF transformations, as defined in Ref.~\cite{QRF}.

Finally, we find that this description of QRFs in spacetime allows us to observe a \emph{superposition of special relativistic time dilations} and a \emph{superposition of gravitational redshifts}, which could be measured experimentally in the future.

Overall, our model paves the way for a full extension of the formalism of QRFs to arbitrary spacetimes, and superpositions thereof. This is an important step to achieve a fully relational formulation of physics on a nonclassical spacetime, which is a key requirement to formulate a quantum theory of gravity.

The structure of the paper is as follows. In Section~\ref{sec:Formalism}, we introduce the model for spacetime QRFs: we start from a fully constrained description of the system, which is neutral to the specific perspective, and is encoded in a set of first-class constraints. We then show how to recover a relational description from the perspective of one of the particles. In Section~\ref{sec:ExtSym}, we show that this construction leads to an extension of the spacetime symmetries to the set of QRF transformations. This result generalises to the spacetime picture and to quantum clocks the notion of extended symmetry transformation introduced in Ref.~\cite{QRF}.
In Section~\ref{sec:GravRed}, we introduce a measurement model and show how the phenomenon of \emph{superposition of special relativistic time dilation} and \emph{superposition of gravitational redshift arises}. For clarity, the technicalities are kept to the essential in the main text, however the relevant calculations are detailed in the Appendices, which are referred to where appropriate.

\section{From a covariant timeless model to relational dynamics in a locally inertial quantum reference frame}
\label{sec:Formalism}

\begin{figure}[t]
	\begin{center}
		\includegraphics[scale=0.35]{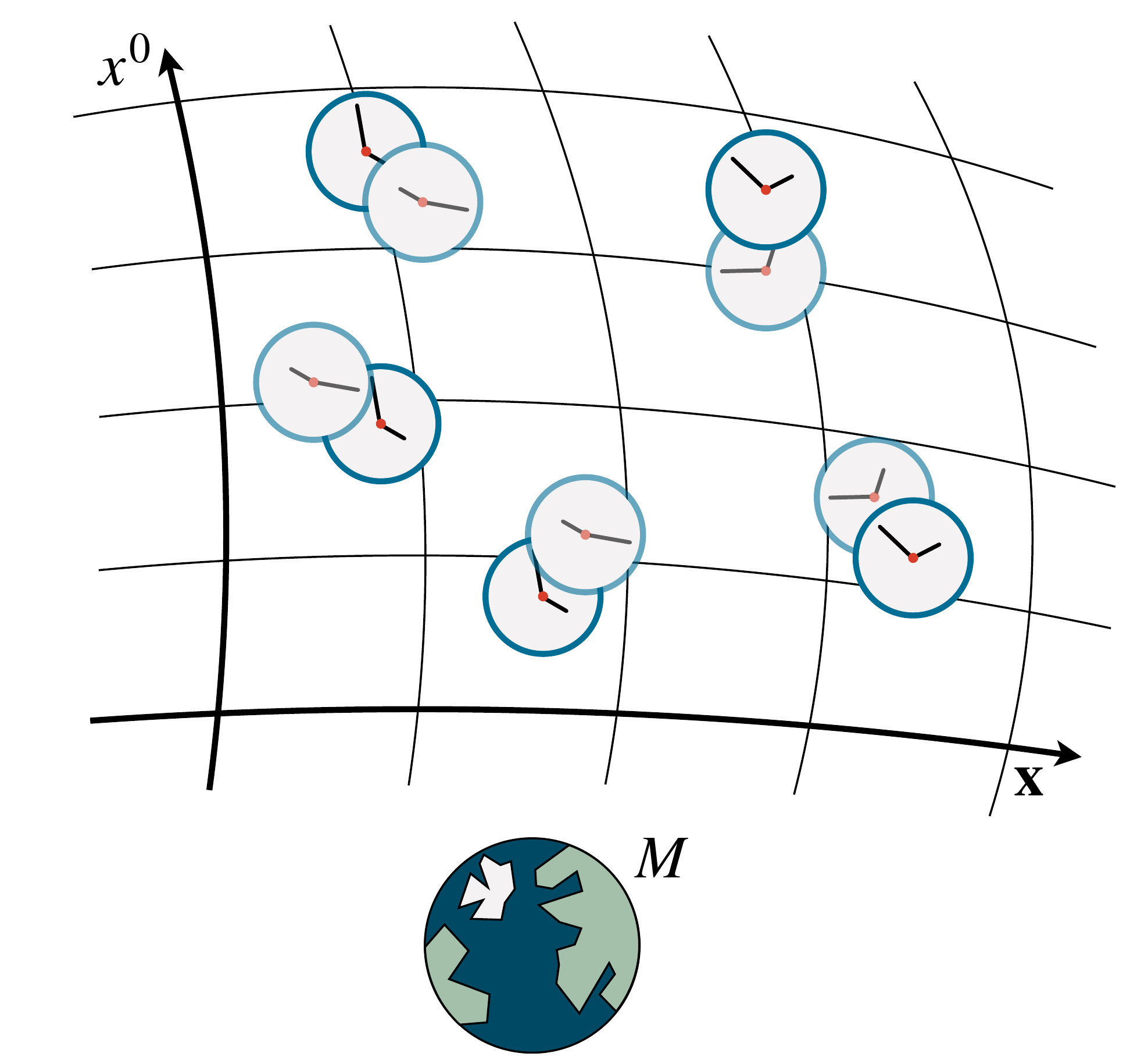}
		\caption{\label{fig:GlobalModel} We consider a system of $N$ relativistic quantum particles in a weak gravitational field produced by a mass $M$. Each particle has a quantum state representing its position in the spacetime diagram, as well as a clock state (the hands of the clocks) which keeps the proper time in the particle's frame. We introduce a timeless model, in which the global state of the $N$ particles and of the mass $M$ is ``frozen'', and is described by an $N$-particle quantum state in spacetime. Concretely, this means that the quantum state of the external variables can be, for instance, in a quantum superposition of coordinate times $x^0$ and spatial coordinates $\mathbf{x}$. Classically, the position in spacetime and the velocity of a particle influence its proper time due to relativistic time dilation. When the particle is in a quantum superposition of positions or velocities, the proper time displayed by its internal clock is also in a quantum superposition from an external perspective. \\
		While, at the global level, the system does not evolve, the dynamical evolution is recovered in terms of the relational variables between one of the particles, which is chosen as the quantum reference frame (QRF), and the rest of the particles. To transform to the QRF of one of the particles, we first map the initial spacetime coordinates to the relational spacetime coordinates from the perspective of the particle chosen as the QRF. We then find a transformation which makes the metric locally inertial at the origin of the QRF. Finally, we use the proper time of the particle to parametrise the dynamics of the remaining particles from the perspective of the chosen QRF. In such QRF, the dynamical evolution of the remaining particles  is described in terms of a Hamiltonian operator and is parametrised by the clock's proper time, which is just a classical parameter in the clock's rest frame.}
	\end{center}
\end{figure}

In this section, we introduce a fully relational model for $N$ quantum particles in spacetime, where
\begin{itemize}
	\item Space and time are treated on equal footing, as in special and general relativity, also when the systems under study are quantum, and not classical;
	\item Time evolution emerges from internal correlations of physical clocks, which are described as living in the internal Hilbert spaces of the particles. The way in which the dynamical laws are recovered from the perspective of such clocks incorporates the Page-Wootters formulation \cite{Page:1983uc, giovannetti2015quantum};
	\item The formalism is fully relational, i.e., all external spacetime structure is eliminated.
\end{itemize}
Furthermore, we show that it is possible to take the perspective of one of these quantum particles, and how the transformation to such QRF is constructed.

We consider a system of N particles of mass $m_I$, for $I=1, \cdots, N$, each of which has a quantum state living in the tensor product of an external Hilbert space, representing its position in spacetime or momentum, and an internal Hilbert space (the clock), and a mass $M$ which is the source of the gravitational field, as illustrated in Fig.~\ref{fig:GlobalModel}. The ideas we present in this paper are described in their simplest form in the time and radial components of a model with spherical symmetry. To simplify the presentation, we work in $1+1$ dimensions, which allows us to neglect the angular part of the model. We adopt a timeless formulation \cite{Page:1983uc, rovelli1990quantum, reisenberger2002spacetime, hellmann2007multiple, giovannetti2015quantum} describing the full state of the particles. The intuitive idea behind these formulations is that the dynamics of particles is an emergent phenomenon stemming from the relational degrees of freedom. Formally, this means that the state of the $N$ particles and of the mass $M$ does not evolve, but the dynamics is recovered by ``conditioning'' on one of the systems considered. The global state of the $N$ particles and of the mass $M$ ---~called the \emph{physical state}~---~is then a ``frozen'' (i.e., non-evolving) state $\ket{\Psi}_{ph}$ satisfying the relation
\begin{equation} \label{eq:SolConst}
	\hat{C}\ket{\Psi}_{ph} =0,
\end{equation}
where $\hat{C}$ is a set of first-class constraints defined on the Hilbert space of the $N$ particles and of the mass $M$. This fully constrained model is not reference-frame specific, but corresponds to a neutral perspective. The set of constraints encodes both the dynamics of the particles in spacetime and the global symmetries (global translations in space and time) of the model. In this Section, we discuss in detail the form of the constraints and their physical meaning.

Differently to other formulations for QRFs, we here describe both the quantum state associated to the position/momentum of the particle and the clock state, and they both play a role in the identification of the QRF. In particular, when one of the particles, say particle $1$, acts as a QRF, the external part of the quantum state is used to fix the QRF (in our case, the origin of the coordinate system in spacetime), while the internal part serves as a clock, ticking according to the proper time in each particle's frame. We will see that imposing the constraints corresponds to removing absolute (or external) variables from the description, and that the relational variables, defined from the point of view of one of the physical systems considered, evolve dynamically in a non-trivial way. In particular, the presence of the internal clocks is crucial to recover the relational dynamics of the rest of the particles from the point of view of the particle chosen as the QRF. 

Throughout the paper, we use the following notation: Greek letters indicate spacetime labels, i.e., $\mu = 0,1$, while capital Latin letters label the particles from $1$ to $N$. Vectors with no indices are, unless differently specified, two-vectors in $1+1$ spacetime, i.e., $v = (v^0, \mathbf{v})$, where $v^0$ is the time component of the vector and $\mathbf{v}$ denotes the spatial component of the vector. We use the Einstein's convention on all sums, unless explicitly stated.

We take the particles to be special-relativistic and in a Newtonian field generated by a mass $M$. We describe the latter as the weak-field limit of a more general gravitational field, corresponding to a metric
\begin{equation} \label{eq:NewtonianG}
	\begin{split}
		& g_{00} = 1 + \frac{2\Phi(\mathbf{x}- \mathbf{x}_M)}{c^2};\\
		& g_{01} = g_{10} = 0;\\
		& g_{11} = -1,
	\end{split}
\end{equation}
where $\Phi(\mathbf{x})$ is the Newtonian potential due to the mass $m_M$ of the system $M$ and $|\Phi(\mathbf{x})|/c^2 \ll 1$ in the spacetime region considered. 

The external Hilbert space of each particle $I=1, \cdots, N$ is $\mathcal{H}_I \simeq L^2 (\mathbb{R}^2)$. The mass $M$ is also assigned a Hilbert space $\mathcal{H}_M \simeq L^2 (\mathbb{R}^2)$. The motion of any of the $N$ particles in this external, weak gravitational field sourced by the mass $M$ can be encoded in the expression
\begin{equation} \label{eq:CIconstraint}
	\hat{C}_I = \sqrt{g^{00}(\hat{\mathbf{x}}_I - \hat{\mathbf{x}}_M)}\hat{p}^I_0 -\hat{\omega}_p^I,
\end{equation}
where $[\hat{x}^\mu_I, \hat{p}_\nu^I] = \mi \hbar \delta^\mu_\nu$, $\hat{\omega}_p^I =  \sqrt{m_I^2 c^2 + \hat{\mathbf{p}}^2_I}$ with $I=1, \cdots, N$, and $[\hat{x}^\mu_M, \hat{p}_\nu^M] = \mi \hbar \delta^\mu_\nu$. The constraints $\hat{C}_I$ of Eq.~\eqref{eq:CIconstraint} straightforwardly follow from the quantisation of the (classical) general-relativistic dispersion relation $g^{\mu\nu}(x) p_\mu p_\nu -m^2 c^2 =0$ thanks to the weak-field approximation, which is crucial to avoid ordering ambiguities. Notice that, for simplicity, we assume that the state of the mass $M$ does not undergo any dynamical evolution in this initial description, hence we do not associate a dynamical constraint to it. In the following we show that the mass acquires a dynamical evolution as a result of taking the perspective of one of the particles.

By applying Eq.~\eqref{eq:CIconstraint} to a quantum state $\ket{\psi_I}_{ph}$ that solves the constraint, i.e., $\hat{C}_I\ket{\psi_I}_{ph} =0$ and conditioning on the time coordinate $\ket{x^0_I}_I$ via the procedure $\braket{x^0_I | \hat{C}_I | \psi_I}_{ph} =0$, it is possible to recover the Schr{\"o}dinger equation for a quantum particle in the Newtonian field (see Appendix~\ref{App:lowenergy} for details). Hence, the dynamical evolution emerges from the entanglement, at the level of $\ket{\psi_I}_{ph}$, between the time and spatial degrees of freedom.

We are working here in a perturbative regime in which the commutator $\left[\sqrt{g_{00}(\hat{\mathbf{x}}_I - \hat{\mathbf{x}}_M)}, \hat{\omega}_p^I \right]$ is negligible. This is equivalent to demanding that all terms which are at least of the order $\frac{\Phi (\hat{\mathbf{x}}_I - \hat{\mathbf{x}}_M)\hat{\mathbf{p}}^2_I}{m_I^2 c^4}$ are negligible whenever they are applied to the quantum state of each particle $I$. Notice that we retain terms of the order $ \left[\hat{\mathbf{p}}_I/(m_I c) \right]^4$, corresponding to special-relativistic corrections in the velocity of the particles. 

The constraint $\hat{C}_I$ encoding the dynamical evolution of the particle, which holds for each particle of the set we consider, is not yet cast in relational terms. The introduction, compared to the standard quantum-mechanical formalism, of the $\hat{x}_I^0$ and $\hat{p}_0^I$ operators, such that $[\hat{x}_I^0, \hat{p}_0^J] = \mi \hbar \delta_I^J$ is necessary to treat space and time on the same footing, as in the case of Covariant Quantum Mechanics~\cite{rovelli1990quantum, reisenberger2002spacetime}. In order to enforce the relational character of the formalism, we need to eliminate the external structure.  In Appendix~\ref{App:RelQM} we review and compare different relational approaches.

In the following, we show how to build a fully relational model for quantum particles in spacetime and in a weak gravitational field. Methodologically, this construction adopts the tools of QRFs, and unifies the framework for spatial QRFs~\cite{QRF, perspective1, perspective2} with time reference frames~\cite{castro2020quantum}, the latter in the non-interacting case. The resulting model utilises techniques of Covariant Quantum Mechanics in order to recover a full spacetime covariance, and incorporates both external and internal degrees of freedom. 

In particular, we consider two constraints on top of the $N$ constraints of Eq.~\eqref{eq:CIconstraint}, one for each spacetime dimension, which encode the conservation of the total energy and the total momentum of our $N$-particle model. Thus, we write 
\begin{equation} \label{eq:Fconstraint}
	\begin{split}
		& \hat{f}^0 = \sum_{I=1}^N \left[\hat{p}^I_0 +  \Delta(\hat{\mathbf{x}}_I - \hat{\mathbf{x}}_M, \hat{\mathbf{p}}_I) \frac{\hat{H}_I}{c}\right] + \hat{p}_0^M;\\
		& \hat{f}^1 = \sum_{I=1}^N \hat{\mathbf{p}}_I + \hat{\mathbf{p}}_M,
	\end{split}
\end{equation}
where  $\Delta(\hat{\mathbf{x}}_I - \hat{\mathbf{x}}_M, \hat{\mathbf{p}}_I) = \sqrt{g_{00}(\hat{\mathbf{x}}_I - \hat{\mathbf{x}}_M)}\left( 1 + \frac{\hat{\mathbf{p}}^2_I}{m_I^2 c^2}\right)^{-1/2}$. The constraint $\hat{f}^1$ corresponds to the global translational invariance required in Refs.~\cite{perspective1, perspective2}. The constraint $\hat{f}^0$ corresponds to the conservation of the total energy. Notice that particle $M$ counts towards the total energy and momentum balance because, although its state does not obey a dynamical constraint, it is a quantum system in its own right, with energy and momentum associated to it. Such quantum state in spacetime should be chosen so that it corresponds to a physical configuration of the mass $M$ which does not dynamically evolve in the initial coordinates. 

The constraint $\hat{f}^0$ is derived by summing all the contributions to the energies (up to a multiplicative factor of $c$) coming from the motion of the particles in spacetime, from the internal Hamiltonian $\hat{H}_I$, living in the Hilbert space $\mathcal{H}_{C_I} \simeq L^2(\mathbb{R})$ of the $N$ particles, and from the system $M$. In particular, the factor in front of the internal energy Hamiltonian corresponds to the relativistic time dilation from the rest frame to an arbitrary frame. The relation between the rest frame and an external frame to the particle is quantum, because the particles have a quantum state associated to the position/momentum, which can be in a quantum superposition of positions and velocities. QRF techniques justify operationally the change between the rest frame of a quantum system and an arbitrary frame~\cite{QRF, giacomini2019relativistic, streiter2020relativistic}. The relation between the internal energies in the rest frame and in an arbitrary frame is encoded in the operator $\Delta(\hat{\mathbf{x}}_I - \hat{\mathbf{x}}_M, \hat{\mathbf{p}}_I)$, which coincides with the worldline element of particle $I$. In Appendix~\ref{App:DeltaOp} we provide an explicit derivation of the operator $\Delta(\hat{\mathbf{x}}_I - \hat{\mathbf{x}}_M, \hat{\mathbf{p}}_I)$.

The two constraints $\hat{f}^0$ and $\hat{f}^1$, being first-class constraints, generate a gauge transformation on the canonically conjugated variables. These constraints have the effect of reducing the number of degrees of freedom of the full system to obtain the dynamics of the relative positions between the particles. In the following we show that, by picking one of the particles as the preferred point of view from which we describe the dynamics of the remaining particles, the resulting dynamics can be seen as the dynamics of the $N-1$ particles and the mass $M$ from the point of view of the chosen particle.

We then see that the introduction of the zero component of the coordinate and momentum operators not only allows us to treat the $N$-particle system in a covariant way, but allows us to generalise the symmetry principle of Refs.~\cite{perspective1, perspective2} to all the spacetime components and to show the connection with the Page-Wootters inspired approach of Ref.~\cite{castro2020quantum}. When only the internal state of the particles is considered, the constraint $\hat{f}^0$ reduces to the Page-Wootters constraint of the clocks. We stress here that the constraint $\hat{f}^0$ can be equivalently seen as arising from a Page-Wootters construction, which emphasises the relational approach to time, or from a relational formulation of physics, which eliminates the background structure by enforcing that the total energy is zero.

As a result of this construction, we have a fully constrained model describing the $N$ particle system, where the full constraint is
\begin{equation} \label{eq:FullC}
	\hat{C} = \sum_{I=1}^N  \mathcal{N}_I \hat{C}_I + z_\mu \hat{f}^\mu,
\end{equation}
where $ \mathcal{N}_I$, with $I=1, \cdots, N$ and $z_\mu$, with $\mu=0,1$ are Lagrange multipliers. Notice that, to our order of approximation, all constraints are first-class, i.e., they at least weakly commute with each other. In the two cases $\left[ \hat{C}_I, \hat{f}^1 \right] = 0$ and $\left[ \hat{f}^0, \hat{f}^1 \right] = 0$, the relation holds exactly. For the commutator $\left[ \hat{C}_I, \hat{f}^0 \right]=0$, the result can be easily obtained by remembering that we are working in a regime where $\left[ g_{00}(\hat{\mathbf{x}}_I - \hat{\mathbf{x}}_M), \hat{\omega}_p^I \right] = 0$ for all $I=1, \cdots, N$. Notice that, thanks to the fact that all the constraints commute, it is possible, as we will show, to straightforwardly build a state that satisfies all of them.

Here, we only consider non-interacting clocks in order to maintain the commutative property of the constraints. Had we allowed for gravitational interaction between the particles, we would have had non-commutative constraints (namely, second-class constraints), which require a different treatment to what is outlined here. We thus leave this question for future work. 

The physical state of Eq.~\eqref{eq:SolConst} with the constraint $\hat{C}$ of Eq.~\eqref{eq:FullC} can equivalently be written as
\begin{equation} \label{eq:PhysState}
	\ket{\Psi}_{ph} \propto \int d^N \mathcal{N} d^2 z e^{\frac{\mi}{\hbar}\mathcal{N}_I \hat{C}_I}e^{\frac{\mi}{\hbar}z_\mu \hat{f}^\mu}\ket{\phi},
\end{equation}
where $\ket{\phi}$ is an arbitrary state that can be expanded in a basis as
\begin{equation}
	\ket{\phi} = \int \Pi_{I} \left[d \mu(x_I) d E_I\right]d^2 x_M \phi(x_1, \cdots, x_N, x_M, E_1, \cdots, E_N)\ket{x_1, \cdots, x_N, x_M}\ket{E_1, \cdots, E_N},
\end{equation}
with $d \mu(x_I) = \sqrt{g_{00}(\mathbf{x}_I-\mathbf{x}_M)} d^2 x_I$ being the covariant integration measure and $\ket{x_I} = \ket{x_I^0, \mathbf{x}_I}$. The states $\ket{E_I}$ are the eigenstates of the internal hamiltonian of each particle $I$, such that $\hat{H}_I \ket{E_I} = E_I \ket{E_I}$. The state $\ket{\phi}$, in general, is not a solution of the constraint. However, the state that solves the constraint $\hat{C}\ket{\Psi}_{ph} = 0$ can be obtained via the improper projection $\hat{P}\ket{\phi} = \ket{\Psi}_{ph}$ defined in Eq.~\eqref{eq:PhysState}, where $\hat{P} = \int d^N \mathcal{N} d^2 z e^{\frac{\mi}{\hbar}\mathcal{N}_I \hat{C}_I}e^{\frac{\mi}{\hbar}z_\mu \hat{f}^\mu}$. The procedure to obtain the state $\ket{\Psi}_{ph}$ involves a technical subtlety, due to the fact that $\hat{P}^2 \neq \hat{P}$. For instance, in situations in which the spectrum of the constraint is continuous around zero, the solution the set of states satisfying the constraint is empty. However, this problem can be solved, e.g., by redefining the inner product in order to renormalise the state $\ket{\Psi}_{ph}$ (for details and different approaches to the solution see, e.g., \cite{ashtekar1991lectures, Marolf1995Refined, Hartle:1997dc, kempf2001implementation}).

In order to take the perspective of a specific particle, e.g., particle $1$, we need to first map the phase-space observables to the relational observables to particle $1$, and then enforce that particle $1$ is in the origin of the reference frame via a projection operation. This procedure extends the technique introduced in Ref.~\cite{perspective1, perspective2} to our model for $N$ relativistic particles in a weak gravitational field. Overall, we obtain the relational state from the perspective of particle $1$ to be
\begin{equation} \label{eq:history1}
	\ket{\psi}^{(1)} = {}_1\bra{q_1=0} \hat{\mathcal{T}}_1 \ket{\Psi}_{ph},
\end{equation}
where we choose
\begin{equation}
	\hat{\mathcal{T}}_1 = e^{\frac{i}{\hbar}\frac{\log\sqrt{g_{00}(\hat{\mathbf{x}}_M)}}{2}\sum_{I=1}^N(\hat{x}^0_I \hat{p}_0^I + \hat{p}_0^I  \hat{x}^0_I)} e^{\frac{\mi}{\hbar}\hat{\mathbf{x}}_1 \left(\hat{f}^1 - \hat{\mathbf{p}}^1 \right)} e^{\frac{\mi}{\hbar}\hat{x}_1^0 \left( \hat{f}^0 - \hat{p}^1_0 \right)},
\end{equation}
and we have defined the positions after the transformation $\hat{\mathcal{T}}_1$ as $q_I$.

The operator $\hat{\mathcal{T}}_1$ maps the initial positions to the relative positions to particle $1$. Its construction can be intuitively explained by distinguishing two parts of the operator. The two rightmost terms map the spacetime coordinates $\hat{x}_\ell$, with $\ell = 2, \cdots, N, M$ to the relative coordinates to particle $1$, i.e., $ \hat{x}_\ell - \hat{x}_1 \mapsto \hat{x}_\ell$. The last term on the left, which is a function of the metric, sets the metric field to a locally inertial metric field at the location of particle $1$. In order to achieve this, the metric at the location of particle $1$ is mapped to the Minkowski metric, and the motion of the other particles is encoded in a new constraint
\begin{equation}
	\hat{C}'_i= \hat{\mathcal{T}}_1 \hat{C}_i \hat{\mathcal{T}}_1^\dagger = \sqrt{\frac{g_{00} (\hat{\mathbf{q}}_i - \hat{\mathbf{q}}_M )}{g_{00} (\hat{\mathbf{q}}_M)}} \hat{k}_0^i - \hat{\omega}_k^i,
\end{equation}
where $i=2, \cdots, N$, $ \hat{q}_i$, $\hat{q}_M$ being the (spacetime) position operators of the particles relative to particle $1$ in the new coordinate system and $\hat{k}_i$, $\hat{k}_M$ the momentum operators that are canonically conjugated respectively to $\hat{q}_i$ and $\hat{q}_M$, i.e., $\left[ \hat{q}_i^\mu, \hat{k}_\nu^j\right] = \mi \hbar \,\delta^\mu_\nu\, \delta_{ij}$ and $\left[ \hat{q}_M^\mu, \hat{k}_\nu^M \right] = \mi \hbar \,\delta^\mu_\nu$ (see Appendix~\ref{App:ActionT} for the complete action of the operator $\hat{\mathcal{T}}_1$ on the phase-space operators). We hence find that
\begin{equation} \label{eq:gQLIF}
	g'_{00} (\hat{\mathbf{q}}_i, \hat{\mathbf{q}}_M ) = \hat{\mathcal{T}}_1 \frac{g_{00} (\hat{\mathbf{x}}_i - \hat{\mathbf{x}}_M )}{g_{00} (\hat{\mathbf{x}}_1 - \hat{\mathbf{x}}_M )} \hat{\mathcal{T}}_1^\dagger,
\end{equation}
is the new metric field in the perspective of particle $1$. In summary, the transformation $\hat{\mathcal{T}}_1$ realises a transformation to the Quantum Locally Inertial Frame (QLIF) centred in particle $1$, in the spirit of Ref.~\cite{giacomini2020einstein}. A different choice of the operator $\hat{\mathcal{T}}_1$ is possible, and would correspond to a different set of relational variables (and coordinate system).

By performing a lengthy calculation (which is detailed in Appendix~\ref{App:calculation}), it is possible to show that the state of Eq.~\eqref{eq:history1} can be cast as
\begin{equation} \label{eq:historyHam1}
	\ket{\psi}^{(1)} = \int d \tau_1 e^{-\frac{\mi}{\hbar} \hat{H}^{(1)} \tau_1} \ket{\psi^{(1)}_0} \ket{\tau_1},
\end{equation}
where $\ket{\tau_1}$ is the state of the internal clock $1$ such that, given the internal time operator $\hat{T}_I$, $\hat{T}_I \ket{\tau_I} = \tau_I \ket{\tau_I}$ and $\left[ \hat{T}_I, \hat{H}_J \right]= \mi \hbar \delta_{IJ}$, the Hamiltonian $\hat{H}^{(1)}$ is
\begin{equation} \label{eq:H1}
	\hat{H}^{(1)}= \hat{\gamma}_{\Sigma p,1} \sum_i \sqrt{g'_{00}(\hat{\mathbf{q}}_i, \hat{\mathbf{q}}_M)}\left\lbrace c\hat{\omega}_k^i + \hat{\gamma}_i^{-1} \hat{H}_i \right\rbrace + c \hat{\gamma}_{\Sigma k,1} \sqrt{g^{00}(\hat{\mathbf{q}}_M)} \hat{k}_0^M + m_1 c^2 \hat{\gamma}_{\Sigma k,1}^2,
\end{equation}
with $\hat{\gamma}_{\Sigma k,1}= \sqrt{1 + \frac{\left(\sum_{i}\hat{\mathbf{k}}_{i} + \hat{\mathbf{k}}_M\right)^2}{m_{1}^2 c^2}}$, $\hat{\gamma}_{i} = \sqrt{1 + \frac{\hat{\mathbf{k}}_{i}^2}{m_{i}^2 c^2}}$. Finally, the state $\ket{\psi^{(1)}_0}$ formally plays the role of an initial state, whose relation with the state $\ket{\phi}$ is given in Appendix~\ref{App:calculation}. The state of Eq.~\eqref{eq:historyHam1} can be interpreted as a ``history state'', namely a state that associates to any arbitrary time eigenstate $\ket{\tau_1}$ read in the QRF of particle $1$ a state $\ket{\psi^{(1)}_{\tau_1}} = e^{-\frac{\mi}{\hbar} \hat{H}^{(1)} \tau_1} \ket{\psi^{(1)}_0}$ of the remaining particles at the time $\tau_1$ read by the clock. A further projection on the state of the clock $1$, i.e., $\braket{\tau_1 | \psi^{(1)}}$, recovers the standard description of quantum mechanics.

From the expression in Eq.~\eqref{eq:H1}, we see that the Hamiltonian of the particles $i= 2, \cdots, N$ is affected by the presence of particle $1$. On the one hand, this is because the metric field at the location of the other particles, in the QLIF of particle $1$, is a quantum operator depending also on the initial value of the metric field at the location of particle $1$, as it is clear by looking at Eq.~\eqref{eq:gQLIF}. This is an expression of the relational character of general relativity, extended to when we consider quantum particles as QRFs: the metric field in each QRF is defined purely in terms of relational quantities (operators) of the particles and the reference frame. On the other hand, there is also a contribution from the special-relativistic time-dilation operator. This contribution is due to the motion of particle $1$ and appears as the operator $\hat{\gamma}_{\Sigma k,1}$. The interpretation and the operational consequences of this result are discussed in Section~\ref{sec:GravRed}. 

In this Section, we have outlined the method to reduce to the perspective of one of the particle in the general model that we have introduced. However, this model has a well-defined limit to the Galilean free-particle case, the special relativistic case, and the Newtonian case of slowly-moving particles. In all these cases, a similar procedure to what is described in this Section can be defined. All these limiting cases are detailed in Appendix~\ref{App:Limits}.

\section{Extended symmetries and QRF changes}
\label{sec:ExtSym}

The construction we have introduced in the previous section is compatible with an extended notion of spacetime symmetries. This means that every quantum particle can equally serve as a QRF, and that no preferred QRF is singled out. This is easy to see by noticing that the procedure we have followed to transform to the QLIF of particle $1$ can be repeated if we take another particle, i.e., particle $2$, as the QRF. In this case, we obtain the same result, but with all indices $1$ and $2$ swapped. The ``history state'' from the point of view of particle $2$ is then
\begin{equation} \label{eq:history3}
	\ket{\psi}^{(2)} = \int d \tau_2 e^{-\frac{\mi}{\hbar} \hat{H}^{(2)} \tau_2} \ket{\psi^{(2)}_0} \ket{\tau_2},
\end{equation}
where all quantities are defined as in the previous section, and
\begin{equation}
	\hat{H}^{(2)}= \hat{\gamma}_{\Sigma u,2} \sum_{k\neq 2} \sqrt{g'_{00}(\hat{\mathbf{r}}_k, \hat{\mathbf{r}}_M)}\left\lbrace c\hat{\omega}_u^k + \hat{\gamma}_u^{-1} \hat{H}_k \right\rbrace + c \hat{\gamma}_{\Sigma u,2} \sqrt{g^{00}(\hat{\mathbf{r}}_M)} \hat{u}_0^M + m_2 c^2 \hat{\gamma}_{\Sigma u,2}^2,
\end{equation}
with $\hat{\gamma}_{\Sigma u,2}= \sqrt{1 + \frac{\left(\sum_{k\neq 2}\hat{\mathbf{u}}_{k} + \hat{\mathbf{u}}_M\right)^2}{m_{2}^2 c^2}}$. Here, $\hat{r}_k$, with $k=1,3,\cdots, N$, and $\hat{r}_M$ are the position operators in the QRF of particle $2$, and $\hat{u}_k$, $\hat{u}_M$ their conjugate momenta.

This construction then induces an invertible transformation from the reduced ``history state'' of Eq.~\eqref{eq:history1} to the reduced ``history state'' of Eq.~\eqref{eq:history3} which amounts to changing the QRF from the particle $1$ to the particle $2$. It is then clear by construction, and can also be checked by direct computation, that the state of Eq.~\eqref{eq:history3} is obtained from Eq.~\eqref{eq:history1} via the operation
\begin{equation}
	\ket{\psi}^{(2)} = {}_2\bra{r_2=0} \hat{\mathcal{T}}_{12} \ket{\psi}^{(1)} \otimes \ket{p_1=0}_1,
\end{equation}
where $\hat{\mathcal{T}}_{12} = \hat{\mathcal{T}}_{2} \hat{\mathcal{T}}_1^\dagger$ (see Appendix~\ref{App:ActionT} for the action of this operator on the phase space operators). This construction parallels the one introduced in Ref.~\cite{perspective1}.

We have thus found that the Hamiltonian from the perspective of particle $1$ is mapped to a Hamiltonian that is form-invariant, but where all labels $1$ and $2$ are swapped, by a reversible QRF transformation. Thus, we have generalised the notion of an extended symmetry introduced in Ref.~\cite{QRF} to the full spacetime picture and quantum relativistic particles in a weak gravitational field.

\section{Gravitational redshift and  relativistic time-dilation from a QRF}
\label{sec:GravRed}

\begin{figure}\centering	 \large
\hfill 
\raisebox{-\height}{\textbf{a)}}
\raisebox{-\height}{\includegraphics[scale=0.4]{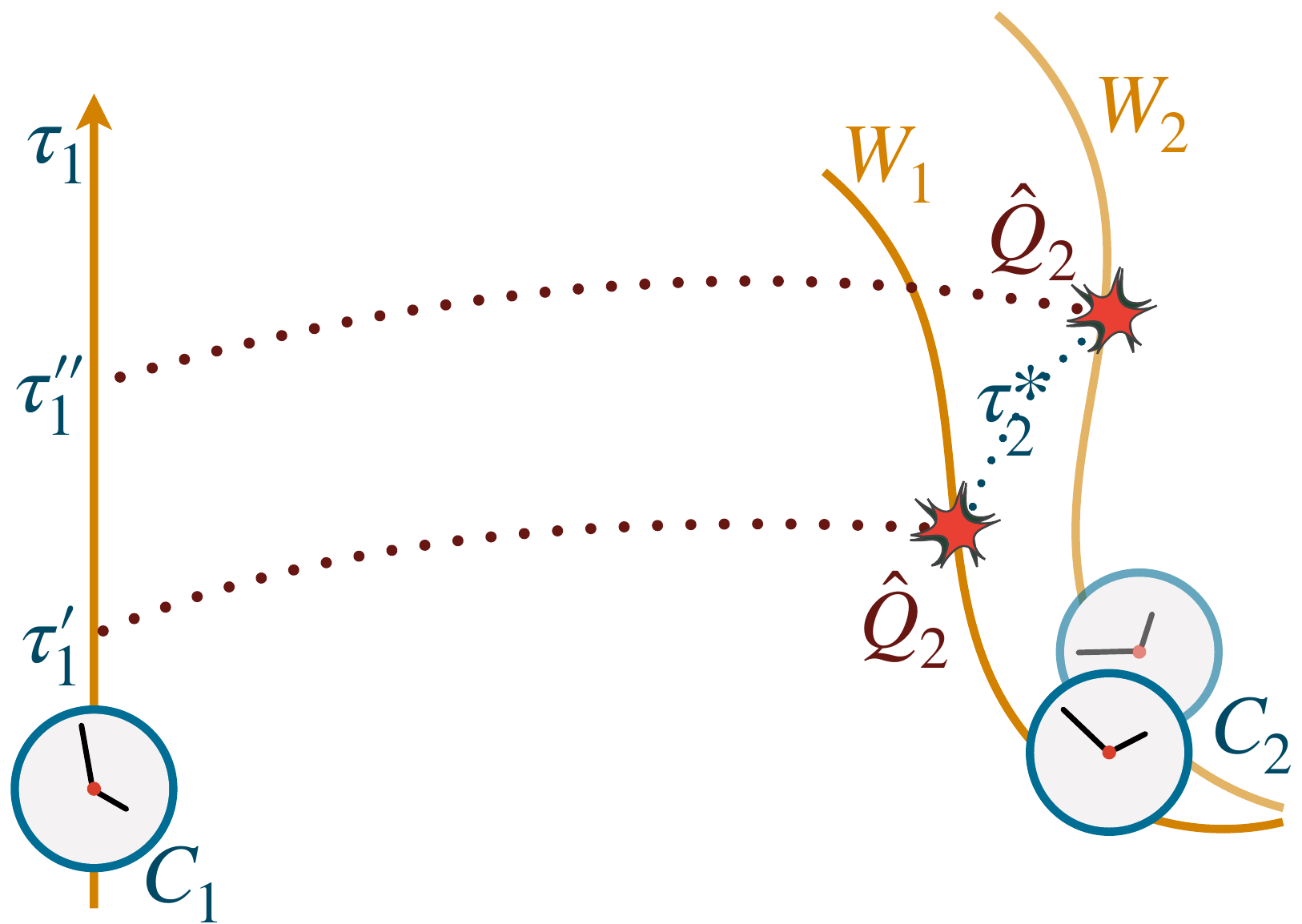}} \hfill
\raisebox{-\height}{\textbf{b)}}
\raisebox{-\height}{\includegraphics[scale=0.4]{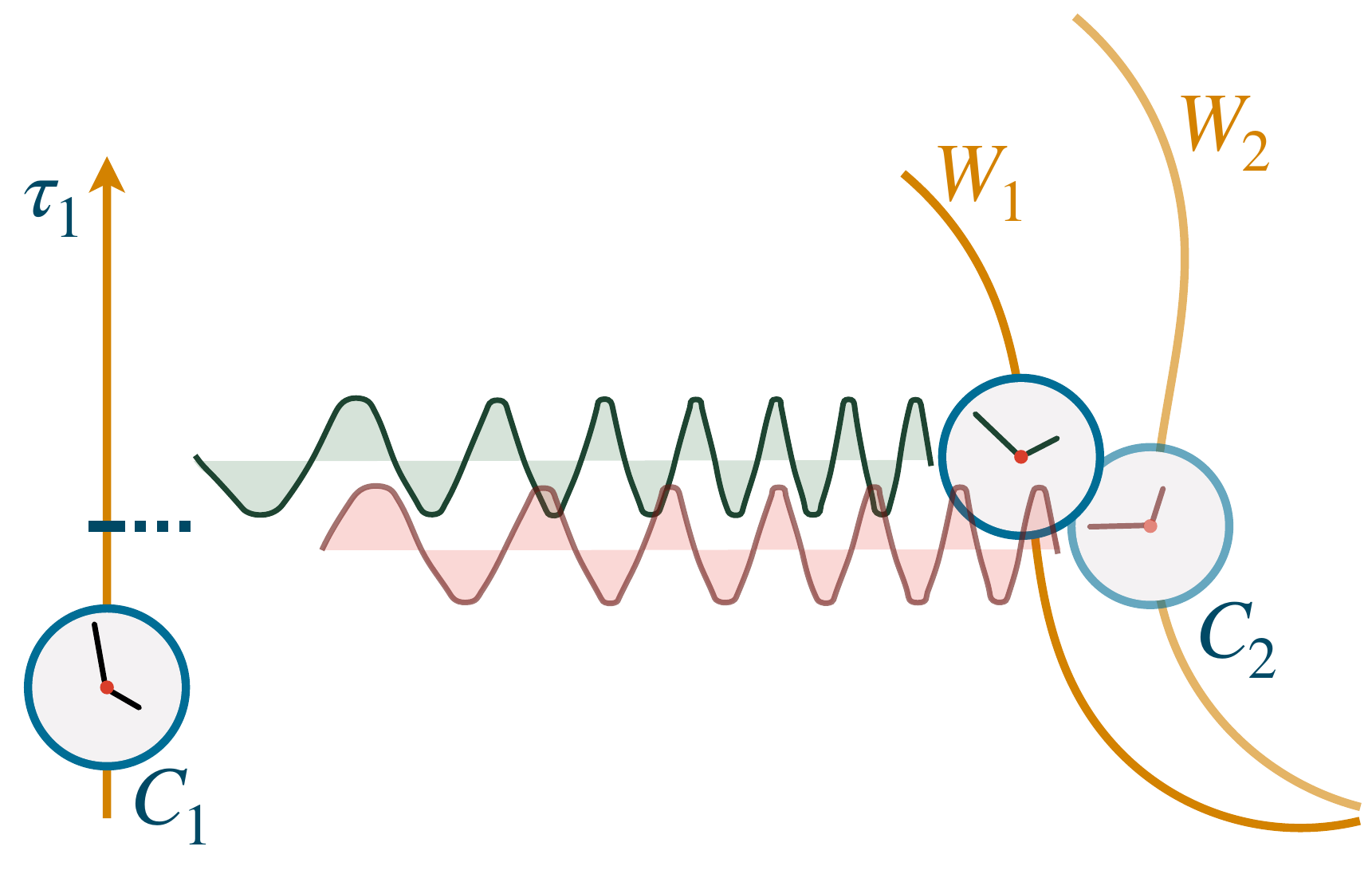}} \hfill\,\\
\caption{\label{fig:SupEffect} We depict the particular situation in which particle $2$ ($C_2$ in the picture) evolves in a superposition of two semi-classical trajectories from the perspective of particle $1$ ($C_1$ in the picture). In the most general case, the state of particle $2$ from the point of view of particle $1$ is arbitrary. We describe a measurement performed at time $\tau_2^*$ in the frame of clock $2$  from the point of view of $C_1$. \textbf{a)} The measurement, which is localised in time in the frame of $C_2$, appears delocalised in time in the frame $C_1$. In particular, it is performed in a superposition of proper times $\tau_1'$ and $\tau_1''$ in the proper time of $C_1$. The two times $\tau_1'$ and $\tau_1''$ are related to $\tau_2^*$ by the expressions $\tau_1' = \Delta_{12}^{-1}(W_1)\tau_2^*$ and $\tau_1'' = \Delta_{12}^{-1}(W_2)\tau_2^*$, where $\Delta_{12}^{-1}(W_i)$, $i=1,2$, encodes the special-relativistic time dilation or the gravitational redshift evaluated on the worldline $W_{i}$. \textbf{b)} In general, $C_1$ ``sees'' the clock $C_2$ as ticking in a superposition of times, depending on the state of the external degrees of freedom. Specifically, this effect can be understood as a superposition of special-relativistic time dilations, when the particles move at relativistic velocities in a Minkowski background, or as a superposition of gravitational redshifts, when the particles move slowly in a weak gravitational field.}
\end{figure}

We now restrict our consideration to $N=2$ particles and a source mass $M$ and introduce the possibility of performing a measurement in the QRF of particle $2$. Analogously to the measurement procedure of Ref.~\cite{castro2020quantum}, we modify the constraint $\hat{f}^0$ as
\begin{equation} \label{eq:CconstraintMeas}
	\begin{split}
		& \hat{C}_I = \sqrt{g^{00}(\hat{\mathbf{x}}_I - \hat{\mathbf{x}}_M)}\hat{p}^I_0 -\hat{\omega}_p^I \qquad \text{for}\qquad I=1, 2;\\
		& \hat{f}_Q^0 = \hat{p}^1_0 + \hat{p}^2_0 + \hat{p}_0^M + \Delta(\hat{\mathbf{x}}_1 - \hat{\mathbf{x}}_M, \hat{\mathbf{p}}_1) \frac{\hat{H}_1}{c} + \Delta(\hat{\mathbf{x}}_2 - \hat{\mathbf{x}}_M, \hat{\mathbf{p}}_2)\left[\frac{\hat{H}_2}{c}+\delta(\hat{T}_2 - \tau_2*) \frac{\hat{Q}_2}{c}\right];\\
		& \hat{f}^1 = \hat{\mathbf{p}}_1 + \hat{\mathbf{p}}_2 + \hat{\mathbf{p}}_M,
	\end{split}
\end{equation}
where $\hat{Q}_2$ is an observable commuting with every constraint\footnote{Notice that this condition can be relaxed by adding an auxiliary quantum state as a measurement device. For the sake of simplicity, we do not consider this case.}. The same procedure adopted in Section~\ref{sec:Formalism} can be used to calculate the history state, with the only \emph{caveat} that the operator $\hat{\mathcal{T}}_1$ needs to be modified to 
\begin{equation}
	\hat{\mathcal{T}}_{Q,1} = e^{\frac{i}{\hbar}\frac{\log\sqrt{g_{00}(\hat{\mathbf{x}}_M)}}{2}\sum_{I=1}^N(\hat{x}^0_I \hat{p}_0^I + \hat{p}_0^I  \hat{x}^0_I)} e^{\frac{\mi}{\hbar}\hat{\mathbf{x}}_1 \left(\hat{f}^1 - \hat{\mathbf{p}}^1 \right)} e^{\frac{\mi}{\hbar}\hat{x}_1^0 \left( \hat{f}_Q^0 - \hat{p}^1_0 \right)}.
\end{equation}
With a similar procedure to the one adopted in Section~\ref{sec:Formalism}, we can calculate the physical state $\ket{\Psi}_{ph}$ and the history state $\ket{\psi}^{(1)} = {}_1\bra{q_1=0}\hat{\mathcal{T}}_{Q,1} \ket{\Psi}_{ph}$. The calculations are detailed in Appendix~\ref{App:CalcMeas}, and the resulting history state is
\begin{equation} \label{eq:historymeas}
	\ket{\psi}^{(1)} = \int d \tau_1 \overleftarrow{T}\left\lbrace  e^{-\frac{\mi}{\hbar}\int_0^{\tau_1} ds \left[\hat{H}^{(1)}+ \Delta_{12} \delta(\hat{T}_2 + \Delta_{12}s - \tau_2^*)\hat{Q}_2 \right]} \right\rbrace \ket{\psi^{(1)}_0} \ket{\tau_1},
\end{equation}
where $\overleftarrow{T}$ denotes the time-ordering operator, defined as $\overleftarrow{T}[\hat{O}(t_1) \hat{O}(t_2)] = \hat{O}(t_1) \hat{O}(t_2)$ if $t_1 > t_2$ and $\overleftarrow{T}[\hat{O}(t_1) \hat{O}(t_2)] = \hat{O}(t_2) \hat{O}(t_2)$ if $t_2 > t_1$ for $\hat{O}(t_1)$, $\hat{O}(t_2)$ being any two arbitrary operators. We have also defined
\begin{equation} \label{eq:Delta12}
	\Delta_{12} = \frac{\Delta(\hat{\mathbf{q}}_2 - \hat{\mathbf{q}}_M, \hat{\mathbf{k}}_2)}{\sqrt{g_{00}(\hat{\mathbf{q}}_M)} \hat{\gamma}_{\Sigma k , 1}^{-1}} = \sqrt{\frac{g_{00}(\hat{\mathbf{q}}_2 - \hat{\mathbf{q}}_M)}{g_{00}(\hat{\mathbf{q}}_M)}}\frac{\hat{\gamma}_2^{-1}}{\hat{\gamma}_{\Sigma k , 1}^{-1}},
\end{equation}
as the ``worldline operator'' of particle $2$ in the perspective of clock $1$. Notice that the metric field in Eq.~\eqref{eq:Delta12} coincides with the metric $g'_{00}(\hat{\mathbf{q}}_2, \hat{\mathbf{q}}_M)$ in the QLIF of particle $1$, which was derived in Section~\ref{sec:Formalism}. The Hamiltonian $\hat{H}^{(1)}$ was defined in Eq.~\eqref{eq:H1}, and $\ket{\psi^{(1)}_0}$ is explicitly calculated in Appendix~\ref{App:CalcMeas}.

In order to understand how a measurement happening at time $\tau_2^*$ in the frame of particle $2$ is seen in the QLIF of clock $1$, we restrict our attention to the Galilean, special relativistic, and Newtonian cases, detailed in Appendix~\ref{App:Limits}. We want to study the case in which the clock $2$ is an ideal clock, which corresponds to the requirement that the initial state $\ket{\psi_0^{(1)}}$ is sharp in the clock $2$ time variable. Mathematically, this condition can be expressed as $\psi_0^{(1)}(\mathbf{p}_2, p_M, t_2) \propto \delta(t_2 - t_2^*) \psi_0^{(1)}(\mathbf{p}_2, p_M)$. In the following, we choose $t_2^* = 0$, equivalent to the initial synchronisation of clocks $1$ and $2$. Our results are summarised in Fig.~\ref{fig:SupEffect}.

In the Galilean case of Appendix~\ref{SubApp:LimitsGalilean}, there is no mass $M$ sourcing the gravitational field, and the special relativistic effects are not present. Hence, $\Delta_{12}$ reduces to the identity operator. We can rewrite the state of Eq.~\eqref{eq:historymeas} as
\begin{equation} \label{eq:historymeasGal}
	\ket{\psi}^{(1)} = \int_{-\infty}^{\tau_2^*} d\tau_1  e^{-\frac{\mi}{\hbar}\hat{H}^{(1)}\tau_1}\ket{\psi^{(1)}_0}\ket{\tau_1}+ \int^{\infty}_{\tau_2^*} d\tau_1  e^{-\frac{\mi}{\hbar}\hat{H}^{(1)}(\tau_1- \tau_2^*)} e^{-\frac{\mi}{\hbar}\hat{Q}_2} e^{-\frac{\mi}{\hbar}\hat{H}^{(1)}\tau_2^*}\ket{\psi^{(1)}_0}\ket{\tau_1}.
\end{equation}
In this case we see that, in the QRF of clock $1$, the measurement happens at the same time $\tau_2^*$ as it would in the QRF of clock $2$.

Let us consider now the special-relativistic case described in Appendix~\ref{SubApp:LimitsSpecRel}. In this case, we take into account the special relativistic effects, but we do not have a mass $M$ sourcing the gravitational field. The operator $\Delta_{12}$ reduces to $\Delta_{12}= \hat{\gamma}_{\Sigma k , 1}/\hat{\gamma}_2 \approx \sqrt{1 + \frac{\hat{\mathbf{k}}_2^2}{m_1^2 c^2} - \frac{\hat{\mathbf{k}}_2^2}{m_2^2 c^2} }$. 

The history state is
\begin{equation} \label{eq:historymeasSR}
	\begin{split}
		\ket{\psi}^{(1)} = \int d \mathbf{k}_2 dt_2 &\left\lbrace \int_{-\infty}^{\tau_k} d\tau_1  e^{-\frac{\mi}{\hbar}\hat{H}^{(1)}\tau_1}\psi_0^{(1)} (\mathbf{k}_2, t_2) \ket{\mathbf{k}_2, t_2}+ \right.\\
		+ &\left. \int^{\infty}_{\tau_k} d\tau_1  e^{-\frac{\mi}{\hbar}\hat{H}^{(1)}(\tau_1- \tau_k)} e^{-\frac{\mi}{\hbar}\hat{Q}_2} \psi_{\tau_k}^{(1)} (\mathbf{k}_2, t_2)\ket{\mathbf{k}_2, t_2}\right\rbrace \ket{\tau_1},
	\end{split}
\end{equation}
where $\ket{\psi_{\tau_k}^{(1)}} = \int dt_2 e^{-\frac{\mi}{\hbar}\hat{H}^{(1)}\tau_k}\psi_0^{(1)} (\mathbf{k}_2, t_2) \ket{\mathbf{k}_2, t_2}$ and $\tau_k = \Delta_{12}^{-1}(\mathbf{k}_2) \tau_2^*$.

From the previous equation we see that the measurement happening at time $\tau_2^*$ in the QRF of clock $2$ is seen by clock $1$ in a quantum superposition of special-relativistic time dilations. The relation to the usual time-dilation can easily be understood by comparison with the classical case by considering two particles, $1$ and $2$, which measure the same interval in coordinate time with their own proper time. In this case, the relation $dx^0 = d\tau_1 \gamma_1 = d \tau_2 \gamma_2$ is precisely the classical equivalent of what we found when we consider two clocks $1$ and $2$ in their own QRF.

Here, for each value $\mathbf{k}_2$ of the momentum of particle $2$, there is a well-defined value of the time dilation, which coincides with the classical one. However, since the relational degrees of freedom between the QRF and particle $2$ are in a quantum superposition of momentum eigenstates, we obtain a coherent superposition of special-relativistic time dilations, which implies that clock $1$ sees the measurement as a delocalised event in time.

These results are in agreement with those presented in Refs.~\cite{smith2019quantizing, smith2020quantum} in the case of a quantum relativistic particle with a sharp momentum, or in a superposition of two momenta\footnote{Formally, each sharp momentum state $\ket{p^*}$ is described in Refs.~\cite{smith2019quantizing, smith2020quantum} by a gaussian wavepacket centred in the momentum $p^*$, and following a semi-classical trajectory.}, and generalise them to arbitrary quantum states of the particles.

Finally, we consider the Newtonian case described in Appendix~\ref{SubApp:LimitsNewtonian}. In this case, we take into account the gravitational effects due to the presence of the mass $M$, but we do not consider the special-relativistic effects. Hence, $\Delta_{12} = \sqrt{g'_{00}(\hat{\mathbf{q}}_2, \hat{\mathbf{q}}_M)} = \sqrt{\frac{g_{00}(\hat{\mathbf{q}}_2 - \hat{\mathbf{q}}_M)}{g_{00}(\hat{\mathbf{q}}_M)}}$ and the history state can be written as
\begin{equation} \label{eq:historymeasNewt}
	\begin{split}
		\ket{\psi}^{(1)} = \int d\mathbf{q}_2 d^2 q_M dt_2 &\left\lbrace \int_{-\infty}^{\tau_q} d\tau_1  e^{-\frac{\mi}{\hbar}\hat{H}^{(1)}\tau_1}\psi_0^{(1)} (\mathbf{q}_2, q_M, t_2) \ket{\mathbf{q}_2, q_M, t_2}+ \right.\\
		+ &\left. \int^{\infty}_{\tau_q} d\tau_1  e^{-\frac{\mi}{\hbar}\hat{H}^{(1)}(\tau_1- \tau_q)} e^{-\frac{\mi}{\hbar}\hat{Q}_2} \psi_{\tau_q}^{(1)} (\mathbf{q}_2, q_M, t_2)\ket{\mathbf{q}_2, q_M, t_2}\right\rbrace \ket{\tau_1},
	\end{split}
\end{equation}
where $\ket{\psi_{\tau_q}^{(1)}} = \int dt_2 e^{-\frac{\mi}{\hbar}\hat{H}^{(1)}\tau_q}\psi_0^{(1)} (\mathbf{q}_2, q_M t_2) \ket{\mathbf{q}_2, q_M, t_2}$ and $\tau_q = \Delta_{12}^{-1}(\mathbf{q}_2, \mathbf{q}_M) \tau_2^*$.

Analogously to the special-relativistic case, we have that the measurement happening at time $\tau_2^*$ in the QRF of clock $2$ is delocalised in the proper time of clock $1$. However, the effect in this case is a quantum superposition of gravitational redshifts, as can be easily seen by recalling the standard expression of the gravitational redshift $\tau_{obs} = \sqrt{\frac{g_{00}(x_{obs})}{g_{00}(x_{em})}}\tau_{em}$, where $\tau_{obs}$ and $x_{obs}$ are respectively the proper time and spacetime coordinates of the observer, and $\tau_{em}$ and $x_{em}$ are respectively the proper time and spacetime coordinates of the emitter. If we identify the observer with clock $1$ and the emitter with clock $2$, we recover the relation $\tau_q = \Delta_{12}^{-1}(\mathbf{q}_2, \mathbf{q}_M) \tau_2^*$. Interestingly, this result can be equivalently obtained in the initial set of coordinates, where $\Delta_{12}  = \sqrt{\frac{g_{00}(\hat{\mathbf{x}}_2 - \hat{\mathbf{x}}_M)}{g_{00}(\hat{\mathbf{x}}_1 - \hat{\mathbf{x}}_M)}}$, or knowing that it is always possible to transform to a QLIF where the metric is locally inertial at the location of clock $1$. In this case, it is enough to calculate $\Delta_{12}$ using the metric in the QLIF, i.e., $\Delta_{12} = \sqrt{\frac{g_{00}(x_{em})}{g_{00}(x_{obs})}} =  \sqrt{g'_{00}(\hat{\mathbf{q}}_2, \hat{\mathbf{q}}_M)}$, because $g_{00}(x_{obs}) = 1$ in the QLIF.

The two effects discussed here, namely the \emph{superposition of special-relativistic time dilations} and the \emph{superposition of gravitational redshifts} are, in principle observable. A promising experimental system to carry out tests on the generalisation of these concepts could be atom interferometry~\cite{tino2020testing}.

An interesting consequence of the delocalisation of the measurement that we have described is that, if we fix some time $\tau_1 = \tau_1^*$, clock $1$ in general ``sees'' a superposition of the measurement having taken place or not. Hence, it seems that whether a measurement has occurred or not is a QRF-dependent feature. It is, however, an open question whether this would allow for a change of the causal relations. Previous work on QRFs~\cite{castro2020quantum} and on causal reference frames~\cite{guerin2018observer} suggests that the causal relations are preserved under change of time reference frame, but other approaches~\cite{Hardy:2018kbp} might allow for a change of the causal relations.

The time delocalisation of events has been studied in Ref.~\cite{castro2020quantum}. However, in that case the mechanism causing the delocalisation was the interaction between the internal degrees of freedom of the clocks, or the choice of a clock with an unsharp initial state. Here, the delocalisation happens due to the fact that the external degrees of freedom of the particle are in a quantum state, which is unsharp in the relevant basis (momentum in the special-relativistic case, and position in the Newtonian case). Other authors have also studied, with different techniques to the one that is developed here, the consequences of proper time running in a superposition due to the external degrees of freedom being in a quantum superposition of positions or momenta~\cite{zych_interferometric, pikovski_TimeDilation, pikovski, smith2020quantum, grochowski2020quantum}. However, these results rely on an external and classical reference frame. Here, we generalise these effects to QRFs in a quantum relationship with each other, and show that they are purely due to the relation between the QRF and a physical system that is described.

Furthermore, we show here that regardless of the state of the particle taken as a QRF in the global model, it is always possible to find a QLIF in which, locally, the metric field is flat and the particle is at rest. This result corroborates the generalisation of the Einstein Equivalence Principle of Ref.~\cite{giacomini2020einstein} and provides a concrete model which describes the QRF change including multiple particles. In particular, this model is suitable to describe an experimental test, such as an interferometric experiment~\cite{tino2020precision}, to verify the superposition of gravitational redshifts and the extension of the Einstein Equivalence Principle. It would be interesting to generalise this result to more general metric fields and to a superposition of metric fields as considered in Ref.~\cite{giacomini2020einstein}, where only the dynamics of a single system was considered.

\section{Discussion}

In this work, we have introduced the notion of a spacetime quantum reference frame, i.e., a reference frame in spacetime associated to a quantum system which can be in a superposition, or entangled from the point of view of another quantum system. We have developed a formulation to describe a set of relativistic quantum particles in a weak gravitational field from the perspective of such spacetime quantum reference frame. In order to achieve this, we developed a covariant formulation describing both the quantum state associated to the position/momentum of the particles and its internal state. The former is used to fix the spacetime quantum reference frame, and the latter is used as the internal clock of the quantum reference frame ticking according to its own proper time. This formulation allows us to compare the dynamical evolution in the proper time of different quantum clocks, when the state of their external variables (position or momentum) is in a quantum superposition from the perspective of one of them. We describe such clocks as being attached to a quantum particle evolving with a Hamiltonian operator, and not having a (semi)classical worldline. We find a transformation to the Quantum Locally Inertial Frame of such clocks, introduced in Ref.~\cite{giacomini2020einstein}, and we describe a \emph{superposition of special-relativistic time dilations} and a \emph{superposition of gravitational redshifts} from the perspective of such spacetime quantum reference frames. 

We adopt a timeless formulation, according to which the global state of $N$ particles appears ``frozen'', but the dynamical evolution is recovered in terms of relational quantities of the particles and the quantum reference frame. Formally, we achieve this by imposing a set of constraints on the model, some of which encode the free evolution of the particles and treat space and time on the same footing, and others impose total energy and momentum conservation.

The model that we derive also includes the Page-Wootters mechanism for non-interacting clocks in Refs.~\cite{Page:1983uc, giovannetti2015quantum, castro2020quantum} as a particular case, when the external degrees of freedom are neglected.

This model is suitable to be extended to more general situations. Here, we have only considered a description in $1+1$ dimensions, and a more general model in $3+1$ dimensions will probably have to face technical difficulties such as those solved in Ref.~\cite{perspective2}. For instance, it would be interesting to explore the possibility of imposing as a constraint the total angular momentum tensor in spacetime, i.e., $M_{\mu\nu}$, with $\mu, \nu = 0, \cdots, 3$, rather than just its spatial components, as done in Ref.~\cite{perspective2}.

Another possible generalisation involves the description of (gravitationally) interacting systems. This will likely require a different way of handling the constraints, which would no longer commute (and would thus be second-class constraints). This generalisation, if successful, would allow us to fully incorporate the interacting clocks model, i.e., the description of time reference frames \cite{castro2020quantum} in the spacetime description of quantum reference frames.

In addition, it would be interesting to generalise the model to arbitrary spacetimes and superpositions thereof and build the transformation to a Quantum Locally Inertial Frame as in Ref.~\cite{giacomini2020einstein}, in the general case.

Finally, we notice that the internal energies enter the model presented here in the total energy balance. This means that they are considered as part of the zero component of the total momentum. In other works, for instance in Refs.~\cite{zych_interferometric, pikovski, CastroRuizE2303, smith2020quantum}, the internal energies contribute instead to the total mass, via the relativistic mass-energy equivalence. These two ways of considering the role of the internal energies, as the zero-component of a vector and as a relativistically invariant quantity are, in principle, different. However, the results presented here are in agreement, to our order of approximation, to those one obtains by considering the internal energy as part of the total mass via the mass-energy equivalence. It would be interesting to investigate the connection between these two methods further, in order to gain a deeper insight on the role of the internal degrees of freedom when both quantum and gravitational effects are relevant.

At the interface of quantum theory and gravity, the notion of a classical spacetime is no longer adequate to describe physical phenomena. It is then crucial to generalise the formulation of the laws of physics to situations in which spacetime is non-classical. One of the main goals of the research on quantum reference frames is to generalise the laws of physics to when reference frames are associated to quantum systems. Here, we have taken a step further and we have associated a reference frame in spacetime to a quantum system. Generalisations of this approach to more general scenarios could pave the way for formulating physics on a quantum spacetime.

\acknowledgments{I would like to thank Lin-Qing Chen, Thomas D. Galley, and Lee Smolin for useful discussions at the early stages of this work. I am also grateful to Achim Kempf, Lorenzo Maccone, and Alexander R. H. Smith for helpful comments on the draft. Research at Perimeter Institute is supported in part by the Government of Canada through the Department of Innovation, Science and Economic Development and by the Province of Ontario through the Ministry of Colleges and Universities.}

\appendix

\section{Mass dispersion relation in a weak gravitational field}
\label{App:lowenergy}

In the Newtonian limit, the metric field is described as in Eq.~\eqref{eq:NewtonianG}, i.e.,
\begin{equation}
	\begin{split}
		& g_{00} = 1 + \frac{2\Phi(\mathbf{x}- \mathbf{x}_M)}{c^2};\\
		& g_{01} = g_{10} = 0;\\
		& g_{11} = -1,
	\end{split}
\end{equation}
where $\Phi(\mathbf{x})$ is the Newtonian potential due to the mass $m_M$ of the system $M$ and $|\Phi(\mathbf{x})|/c^2 \ll 1$ in the spacetime region considered. Hence, the general-relativistic dispersion relation $\tilde{C}_I = g^{\mu\nu}p_\mu p_\nu -m_I^2c^2$ of a classical particle $I$ with mass $m_I$ takes the form 
\begin{equation}
	\tilde{C}_I = g^{00}(\mathbf{x}_I- \mathbf{x}_M) p_0^{I2} - \mathbf{p}_I^2 - m_I^2 c^2.
\end{equation}
We impose the positive energy condition by enforcing that $p^I_0 \geq 0$\footnote{Notice that this condition can only be enforced in the weak-field limit of the gravitational field, i.e., when $|\Phi(\mathbf{x})|/c^2 \ll 1$.}. By noting that $\tilde{C}_I$ can be decomposed as
\begin{equation}
	\tilde{C}_I = \left[\sqrt{g^{00}(\mathbf{x}_I- \mathbf{x}_M)}p^I_0 - \sqrt{\mathbf{p}_I^2 + m_I^2 c^2}\right]\left[\sqrt{g^{00}(\mathbf{x}_I- \mathbf{x}_M)}p^I_0 + \sqrt{\mathbf{p}_I^2 + m_I^2 c^2}\right],
\end{equation}
we see that, if the energies are to be positive, it is enough to consider the new dispersion relation
\begin{equation} \label{Appeq:CI}
	C_I = \sqrt{g^{00}(\mathbf{x}_I- \mathbf{x}_M)}p^I_0 - \sqrt{\mathbf{p}_I^2 + m_I^2 c^2}.
\end{equation}
Clearly, this relation only holds in the low-energy regime of the particle, hence it holds perturbatively in $\frac{\mathbf{p}_I^2}{m_I^2c^2}$. It is now straightforward to quantise Eq.~\eqref{Appeq:CI} and obtain the constraint
\begin{equation} \label{Appeq:quantCI}
	\hat{C}_I = \sqrt{g^{00}(\hat{\mathbf{x}}_I- \hat{\mathbf{x}}_M)}\hat{p}^I_0 - \sqrt{\hat{\mathbf{p}}_I^2 + m_I^2 c^2}.
\end{equation}
If we now define $\hat{\omega}^I_p = \sqrt{\hat{\mathbf{p}}_I^2 + m_I^2 c^2}$, we find the constraint $\hat{C}_I$ of Eq.~\eqref{eq:CIconstraint}. Notice that the weak-field approximation $|\Phi(\mathbf{x})|/c^2 \ll 1$ played a crucial role here to avoid ordering ambiguities in the quantisation procedure, as a more general form of the metric field would also have position-dependent spatial components, which, when quantised, do not commute with their conjugate momentum.

 It is easy to show that the expression $\bra{x_I^0} \hat{C}_I \ket{\psi_I}_{ph} =0$, where $\ket{\psi_I}_{ph}$ is the quantum state of particle $I$ solving the constraint $\hat{C}_I$, is equivalent to
\begin{equation}
	\mi \hbar \frac{d}{d x_I^0} \ket{\psi_I(x^0_I)} = \sqrt{g_{00}(\hat{\mathbf{x}}_I - \hat{\mathbf{x}}_M)}\hat{\omega}_p^I \ket{\psi_I(x^0_I)},
\end{equation} 
where $\ket{\psi_I(x^0_I)} = \braket{x_I^0 | \psi_I}_{ph}$. By defining $x^0_I = ct_I$ and expanding perturbatively, we find
\begin{equation}
	\mi \hbar \frac{d}{d t_I} \ket{\psi_I(t_I)} = \left[m_I c^2 +  \frac{\hat{\mathbf{p}}_I^2}{2m_I} - \frac{\hat{\mathbf{p}}_I^4}{8 m_I^3 c^2} + m_I \Phi(\hat{\mathbf{x}}_I - \hat{\mathbf{x}}_M)\right] \ket{\psi_I(t_I)},
\end{equation}
where we do not consider higher-order terms. The equation above corresponds to Schr{\"o}dinger equation of a particle in a Newtonian field, with first-order special-relativistic corrections.

\section{Comparison between relational approaches}
\label{App:RelQM}

In the literature, there are several, and in principle different, ways of eliminating an external, absolute structure. For instance, the Page-Wootters mechanism~\cite{Page:1983uc, giovannetti2015quantum} considers a set of clocks which, at the global level, satisfy a Hamiltonian constraint as in Eq.~\eqref{eq:SolConst}. Upon conditioning on the state of one of the clocks, one recovers the dynamics of the other clocks. This is usually interpreted as the internal perspective of the clock on which one conditions. The Page-Wootters mechanism was applied to interacting clocks in Refs.~\cite{smith2019quantizing, castro2020quantum}. In particular, in the case of gravitationally interacting clocks studied in Ref.~\cite{castro2020quantum} it was shown that the interaction term between the internal degrees of freedom of the clocks leads to a relative temporal localisation of events, due to the fact that different internal energy states contribute differently to the total mass of the system. In addition, the limits to the measurability of time studied in Ref.~\cite{CastroRuizE2303} could be recovered. In a different approach to relational dynamics, it was shown in Refs.~\cite{perspective1, perspective2} that the formalism for QRFs introduced in Ref.~\cite{QRF} can be derived by imposing a symmetry principle on a set of $N$ particles, and specifically by imposing invariance under global translations and global rotations. This symmetry principle is reminiscent of the construction of Shape Dynamics~\cite{mercati2018shape}, which also inspired the construction of a classical model in which a gravitational arrow of time emerges~\cite{barbour2014identification}. The above-mentioned techniques have been derived in different frameworks and are, in principle, different. However, a connection between different approaches to relational dynamics can be found, as shown in Refs.~\cite{hoehn2019trinity, hoehn2020equivalence} for the case of the Page-Wootters mechanism, relational Dirac observables, and quantum deparametrisation.

\section{Worldline operator of a quantum relativistic particle in a weak gravitational field}
\label{App:DeltaOp}

Let us consider a classical relativistic particle of mass $m$ in a weak gravitational field. The classical equivalent of the constraint considered in the main text (up to factors of order $O(c^{-2})$, is
\begin{equation}
	C_I = \sqrt{g^{00}(\mathbf{x}_I)}p_0^I - m_I c\sqrt{1+ \frac{\mathbf{p}_I^2}{m^2_I c^2}},
\end{equation}
where $\mathbf{p}_I^2$ is the spatial norm of the momentum. Via the Hamilton's equations of motion we have $\dot{x}_I^\mu = \frac{dx_I^\mu}{d\tau} = \frac{\partial C_I}{\partial p^I_\mu}$, which yields
\begin{equation}
	\begin{split}
		&\dot{x}_I^0 = \sqrt{g^{00}(\mathbf{x}_I)},\\
		&\dot{\mathbf{x}}_I = \frac{\mathbf{p}_I}{m_I c}\left( 1 + \frac{\mathbf{p}_I^2}{m_I^2c^2} \right)^{-1/2}.
	\end{split}
\end{equation}
Hence, the line element is, to order $O(c^{-2})$,
\begin{equation}
	\begin{split}
		ds &= \sqrt{g_{00}(\mathbf{x}_I) - \frac{1}{c^2} \left(\frac{d \mathbf{x}_I}{d\tau}\right)^2 \left(\frac{d \tau}{d x^0_I}\right)^2}d x^0_I=\\
		&= \sqrt{g_{00}(\mathbf{x}_I)}\sqrt{1+ \frac{\mathbf{p_I}^2}{m^2_I c^2}}dx_I^0.
	\end{split}
\end{equation}
The quantisation of this expression is straightforward, given that in the regime we consider we have $\left[g_{00} (\hat{\mathbf{x}}), \sqrt{1+ \frac{\hat{\mathbf{p}}_I^2}{m_I^2c^2}}\right] =0$. We then write the ``worldline operator'' as
\begin{equation}
		\Delta (\hat{\mathbf{x}}_I, \hat{\mathbf{p}}_I) = \sqrt{g_{00}(\mathbf{\hat{x}}_I)}\sqrt{1+ \frac{\hat{\mathbf{p}}_I^2}{m_I^2c^2}}.
\end{equation}

\section{Action of the $\mathcal{\hat{T}}_1$ operator on the position and momentum operators}
\label{App:ActionT}

In this Appendix, we calculate the action of the operator $\mathcal{\hat{T}}_1$ on the phase space operators of the quantum particle in spacetime. We have
	\begin{equation*}
		\begin{split}
			&\mathcal{\hat{T}}_1 \hat{x}_1^0 \mathcal{\hat{T}}_1^\dagger = \sqrt{g^{00}(\hat{\mathbf{q}}_M)} \hat{q}_1^0,\hspace{10000pt minus 1fil} \\
			&\mathcal{\hat{T}}_1 \hat{\mathbf{x}}_1 \mathcal{\hat{T}}_1^\dagger = \hat{\mathbf{q}}_1,\\
			&\mathcal{\hat{T}}_1 \hat{x}_i^0 \mathcal{\hat{T}}_1^\dagger = \sqrt{g^{00}(\hat{\mathbf{q}}_M)}\left(\hat{q}_i^0 + \hat{q}_1^0 \right),\\
			&\mathcal{\hat{T}}_1 \hat{\mathbf{x}}_i \mathcal{\hat{T}}_1^\dagger = \hat{\mathbf{q}}_i + \hat{\mathbf{q}}_1, \\
			&\mathcal{\hat{T}}_1 \hat{x}_M^0 \mathcal{\hat{T}}_1^\dagger = \hat{q}_M^0 + \sqrt{g^{00}(\hat{\mathbf{q}}_M)} \hat{q}_1^0,\\
			&\mathcal{\hat{T}}_1 \hat{\mathbf{x}}_M \mathcal{\hat{T}}_1^\dagger = \hat{\mathbf{q}}_M + \hat{\mathbf{q}}_1,
		\end{split}
\end{equation*}
\begin{equation}
	\begin{split}
		&\mathcal{\hat{T}}_1 \hat{p}^1_0 \mathcal{\hat{T}}_1^\dagger = \sqrt{g_{00}(\hat{\mathbf{q}}_M)}\Big(\hat{k}^1_0 - \sum_{i\neq 1} \hat{k}_0^i\Big) - \sum_{i\neq 1} \Delta(\hat{\mathbf{q}}_i - \hat{\mathbf{q}}_M, \hat{\mathbf{k}}_i)\frac{\hat{H}_i}{c}-\hat{k}_0^M  - \Delta\Big(\hat{\mathbf{q}}_M, \hat{\mathbf{k}}_1 - \sum_{i\neq 1} \hat{\mathbf{k}}_i - \hat{\mathbf{k}}_M \Big)\frac{\hat{H}_1}{c}, \\
		&\mathcal{\hat{T}}_1 \hat{\mathbf{p}}_1 \mathcal{\hat{T}}_1^\dagger =\hat{\mathbf{k}}_1 - \sum_{i\neq 1} \hat{\mathbf{k}}_i - \hat{\mathbf{k}}_M,\\
		&\mathcal{\hat{T}}_1 \hat{p}^i_0 \mathcal{\hat{T}}_1^\dagger = \sqrt{g_{00}(\hat{\mathbf{q}}_M)} \hat{k}^i_0,\\
		&\mathcal{\hat{T}}_1 \hat{\mathbf{p}}_i \mathcal{\hat{T}}_1^\dagger =\hat{\mathbf{k}}_i,\\
		&\mathcal{\hat{T}}_1 \hat{p}^M_0 \mathcal{\hat{T}}_1^\dagger = \hat{k}^M_0,\\
		&\mathcal{\hat{T}}_1 \hat{\mathbf{p}}_M \mathcal{\hat{T}}_1^\dagger =\hat{\mathbf{k}}_M,
	\end{split}
\end{equation}
where $i = 2, \cdots, N$.

We can also define a new transformation $\mathcal{\hat{T}}_2$ to the relational variables of particle $2$ by swapping the labels $1$ and $2$ in the transformation $\mathcal{\hat{T}}_1$. We define the relational phase-space operators to particle $2$ as $\hat{r}_j$ (position operator) and $\hat{u}_j$ (momentum operator), for $j=1,3,\cdots, N, M$, and we obtain

\begin{equation*}
	\begin{split}
		&\mathcal{\hat{T}}_2 \hat{x}_2^0 \mathcal{\hat{T}}_2^\dagger = \sqrt{g^{00}(\hat{\mathbf{r}}_M)} \hat{r}_2^{0}, \hspace{10000pt minus 1fil}\\
		&\mathcal{\hat{T}}_2 \hat{\mathbf{x}}_2 \mathcal{\hat{T}}_2^\dagger = \hat{\mathbf{r}}_2,\\
		&\mathcal{\hat{T}}_2 \hat{x}_j^0 \mathcal{\hat{T}}_2^\dagger = \sqrt{g^{00}(\hat{\mathbf{r}}_M)} \left(\hat{r}_j^{0} + \hat{r}_2^{0} \right),\\
		&\mathcal{\hat{T}}_2 \hat{\mathbf{x}}_j \mathcal{\hat{T}}_2^\dagger = \hat{\mathbf{r}}_j + \hat{\mathbf{r}}_2, \\
		&\mathcal{\hat{T}}_2 \hat{x}_M^0 \mathcal{\hat{T}}_2^\dagger = \hat{r}_M^{0} + \sqrt{g^{00}(\hat{\mathbf{r}}_M)}\hat{r}_2^{0},\\
		&\mathcal{\hat{T}}_2 \hat{\mathbf{x}}_M \mathcal{\hat{T}}_2^\dagger = \hat{\mathbf{r}}_M + \hat{\mathbf{r}}_2, \\
	\end{split}
\end{equation*}
\begin{equation}
	\begin{split}		
		&\mathcal{\hat{T}}_2 \hat{p}^2_0 \mathcal{\hat{T}}_2^\dagger = \sqrt{g_{00}(\hat{\mathbf{r}}_M)}\Big(\hat{u}^2_0 - \sum_{j\neq 2}\hat{u}^j_0 \Big) - \sum_{j\neq 2}\Delta(\hat{\mathbf{r}}_j - \hat{\mathbf{r}}_M, \hat{\mathbf{u}}_j)\frac{\hat{H}_j}{c} -\hat{u}_0^M  - \Delta \Big(\hat{\mathbf{r}}_M, \hat{\mathbf{u}}_2 - \sum_{j\neq 2} \hat{\mathbf{u}}_j - \hat{\mathbf{u}}_M \Big)\frac{\hat{H}_1}{c} \\
		&\mathcal{\hat{T}}_2 \hat{\mathbf{p}}_2 \mathcal{\hat{T}}_2^\dagger =\hat{\mathbf{u}}_2 - \sum_{j\neq 2} \hat{\mathbf{u}}_j - \hat{\mathbf{u}}_M\\
		&\mathcal{\hat{T}}_2 \hat{p}^j_0 \mathcal{\hat{T}}_2^\dagger = \sqrt{g_{00}(\hat{\mathbf{r}}_M)} \hat{u}^j_0\\
		&\mathcal{\hat{T}}_2 \hat{\mathbf{p}}_j \mathcal{\hat{T}}_2^\dagger =\hat{\mathbf{u}}_j\\
		&\mathcal{\hat{T}}_2 \hat{p}^M_0 \mathcal{\hat{T}}_2^\dagger = \hat{u}^M_0\\
		&\mathcal{\hat{T}}_2 \hat{\mathbf{p}}_M \mathcal{\hat{T}}_2^\dagger =\hat{\mathbf{u}}_M,
	\end{split}
\end{equation}
where $j = 1, 3, \cdots, N$.

We can also transform from one set of relational phase-space operators to another, e.g., from particle $1$ to particle $2$. In order to do this, we define $\mathcal{\hat{T}}_{12} = \mathcal{\hat{T}}_2 \mathcal{\hat{T}}_1^\dagger$ and find
\begin{equation*}
	\begin{split}
		&\mathcal{\hat{T}}_{12} \hat{q}_1^0 \mathcal{\hat{T}}_{12}^\dagger = \sqrt{\frac{g_{00}(\hat{\mathbf{r}}_1 - \hat{\mathbf{r}}_M)}{g_{00}(\hat{\mathbf{r}}_M)}}\left( \hat{r}_1^0 + \hat{r}_2^0\right), \hspace{10000pt minus 1fil}\\
		&\mathcal{\hat{T}}_{12} \hat{\mathbf{q}}_1 \mathcal{\hat{T}}_{12}^\dagger = \hat{\mathbf{r}}_1 + \hat{\mathbf{r}}_2,\\
		&\mathcal{\hat{T}}_{12} \hat{q}_2^0 \mathcal{\hat{T}}_{12}^\dagger = - \sqrt{\frac{g_{00}(\hat{\mathbf{r}}_1 - \hat{\mathbf{r}}_M)}{g_{00}(\hat{\mathbf{r}}_M)}}\hat{r}_1^0,\\
		&\mathcal{\hat{T}}_{12} \hat{\mathbf{q}}_2 \mathcal{\hat{T}}_{12}^\dagger = -\hat{\mathbf{r}}_1,\\
		&\mathcal{\hat{T}}_{12} \hat{q}_{\ell}^0 \mathcal{\hat{T}}_{12}^\dagger = \sqrt{\frac{g_{00}(\hat{\mathbf{r}}_1 - \hat{\mathbf{r}}_M)}{g_{00}(\hat{\mathbf{r}}_M)}} (\hat{r}_{\ell}^0 - \hat{r}_{1}^0), \\
		&\mathcal{\hat{T}}_{12} \hat{\mathbf{q}}_{\ell} \mathcal{\hat{T}}_{12}^\dagger = \hat{\mathbf{r}}_{\ell} - \hat{\mathbf{r}}_{1}, \\
		&\mathcal{\hat{T}}_{12} \hat{q}_{M}^0 \mathcal{\hat{T}}_{12}^\dagger = \hat{r}_{M}^0 - \sqrt{g^{00}(\hat{\mathbf{r}}_M)}\hat{r}_{1}^0, \\
		&\mathcal{\hat{T}}_{12} \hat{\mathbf{q}}_{M} \mathcal{\hat{T}}_{12}^\dagger = \hat{\mathbf{r}}_{M} - \hat{\mathbf{r}}_{1},
	\end{split}
\end{equation*}
\begin{equation}
	\begin{split}
		&\mathcal{\hat{T}}_{12} \hat{k}^1_0 \mathcal{\hat{T}}_{12}^\dagger = \sqrt{g^{00}(\hat{\mathbf{r}}_1-\hat{\mathbf{r}}_M)}\sqrt{g_{00}(\hat{\mathbf{r}}_M)}\hat{u}^2_0, \hspace{10000pt minus 1fil}\\
		&\mathcal{\hat{T}}_{12} \hat{\mathbf{k}}_1 \mathcal{\hat{T}}_{12}^\dagger =\hat{\mathbf{u}}_2,\\
		&\mathcal{\hat{T}}_{12} \hat{k}^2_0 \mathcal{\hat{T}}_{12}^\dagger = \sqrt{\frac{g^{00}(\hat{\mathbf{r}}_1- \hat{\mathbf{r}}_M)}{g^{00}(\hat{\mathbf{r}}_M)}}\Big( \hat{u}^2_0 - \sum_{j\neq 2} \hat{u}_0^j \Big) -\sqrt{g^{00}(\hat{\mathbf{r}}_1- \hat{\mathbf{r}}_M)} \left[ \sum_{j\neq 2} \Delta_j \frac{\hat{H}_j}{c} + \hat{u}_0^M + \Delta_{\Sigma u,2}\frac{\hat{H}_2}{c}\right], \\
		&\mathcal{\hat{T}}_{12} \hat{\mathbf{k}}_2 \mathcal{\hat{T}}_{12}^\dagger =\hat{\mathbf{u}}_2 - \sum_{j \neq 2}\hat{\mathbf{u}}_j - \hat{\mathbf{u}}_M,\\
		&\mathcal{\hat{T}}_{12} \hat{k}^{\ell}_0 \mathcal{\hat{T}}_{12}^\dagger = \sqrt{\frac{g^{00}(\hat{\mathbf{r}}_1- \hat{\mathbf{r}}_M)}{g^{00}(\hat{\mathbf{r}}_M)}} \hat{u}^{\ell}_0,\\
		&\mathcal{\hat{T}}_{12} \hat{\mathbf{k}}_{\ell} \mathcal{\hat{T}}_{12}^\dagger =\hat{\mathbf{u}}_{\ell},\\
		&\mathcal{\hat{T}}_{12} \hat{k}^{M}_0 \mathcal{\hat{T}}_{12}^\dagger = \hat{u}^{M}_0,\\
		&\mathcal{\hat{T}}_{12} \hat{\mathbf{k}}_{M} \mathcal{\hat{T}}_{12}^\dagger =\hat{\mathbf{u}}_{M},
	\end{split}
\end{equation}
where $\ell= 3, \cdots, N$ and $\Delta_j = \Delta (\hat{\mathbf{r}}_j - \hat{\mathbf{r}}_M, \hat{\mathbf{u}}_j )$ and $\Delta_{\Sigma u, 2} = \Delta \Big(\hat{\mathbf{r}}_M, \hat{\mathbf{u}}_2 - \sum_{j\neq 2}\hat{\mathbf{u}}_j - \hat{\mathbf{u}}_M \Big)$. Notice that the relevant relational operators are those of the particles $2, \cdots, N, M$ from the perspective of particle $1$, which are mapped into quantities which only depend on the operators of particle $2$ via $\hat{u}_2$, which is the transformed constraint, since $\hat{\mathcal{T}}_2 \hat{f}^0 \hat{\mathcal{T}}_2^\dagger = \sqrt{g_{00}(\hat{\mathbf{r}}_M)} \hat{u}^0_2$ and $\hat{\mathcal{T}}_2 \hat{f}^1 \hat{\mathcal{T}}_2^\dagger = \hat{\mathbf{u}}_2$.

\section{Explicit calculation of the ``history state'' in the frame of particle 1}
\label{App:calculation}

We recover in this Appendix the dynamical evolution of the relational degrees of freedom from the state of Eq.~\eqref{eq:PhysState}
\begin{equation} \label{Appeq:PhysState}
	\ket{\Psi}_{ph} \propto \int d^N \mathcal{N} d^2 z e^{\frac{\mi}{\hbar}\mathcal{N}_i \hat{C}_i}e^{\frac{\mi}{\hbar}z_\mu \hat{f}^\mu}\ket{\phi},
\end{equation}
where $\ket{\phi}$ is represented as
\begin{equation}
	\ket{\phi} = \int \Pi_{I} \left[d \mu(x_I) d E_I\right]d^2 x_M \phi(x_1, \cdots, x_N, x_M, E_1, \cdots, E_N)\ket{x_1, \cdots, x_N, x_M}\ket{E_1, \cdots, E_N},
\end{equation}
and $d \mu(x_i) = \sqrt{g_{00}(\mathbf{x}_I-\mathbf{x}_M)} d^2 x_I$ is the covariant integration measure. The set of constraints that we enforce in this model is
\begin{equation}
	\begin{split}
		&\hat{C}_I = \sqrt{g^{00}(\hat{\mathbf{x}}_I - \hat{\mathbf{x}}_M)}\hat{p}^I_0 -\hat{\omega}_p^I, 	\qquad\qquad I=1, \cdots, N\\
		& \hat{f}^0 = \sum_{I=1}^N \left[\hat{p}^I_0 +  \Delta(\hat{\mathbf{x}}_I - \hat{\mathbf{x}}_M, \hat{\mathbf{p}}_I) \frac{\hat{H}_I}{c}\right] + \hat{p}_0^M;\\
		& \hat{f}^1 = \sum_{I=1}^N \hat{\mathbf{p}}_I + \hat{\mathbf{p}}_M,
	\end{split}
\end{equation}
where  $\Delta(\hat{\mathbf{x}}_I - \hat{\mathbf{x}}_M, \hat{\mathbf{p}}_I) = \sqrt{g_{00}(\hat{\mathbf{x}}_I - \hat{\mathbf{x}}_M)}\left( 1 + \frac{\hat{\mathbf{p}}^2_I}{m_I^2 c^2}\right)^{-1/2}$.

In order to find the dynamical evolution, we define an operator $\hat{\mathcal{T}}_1$ which maps the observables to the relational observables from the perspective of particle $1$. We define the operator $\hat{\mathcal{T}}_1$ as
\begin{equation}
	\hat{\mathcal{T}}_1 = e^{\frac{i}{\hbar}\frac{\log\sqrt{g_{00}(\hat{\mathbf{x}}_M)}}{2}\sum_{I=1}^N(\hat{x}^0_I \hat{p}_0^I + \hat{p}_0^I  \hat{x}^0_I)} e^{\frac{\mi}{\hbar}\hat{\mathbf{x}}_1 \left(\hat{f}^1 - \hat{\mathbf{p}}^1 \right)} e^{\frac{\mi}{\hbar}\hat{x}_1^0 \left( \hat{f}^0 - \hat{p}^1_0 \right)}, 
\end{equation}
where $\hat{f}^1 - \hat{\mathbf{p}}^1 = \sum_{i} \hat{\mathbf{p}}_{i} +\hat{\mathbf{p}}_M $, $\hat{f}^0 - \hat{p}^1_0 =  \sum_{i}\left[ \hat{p}_0^{i} + \Delta(\hat{\mathbf{x}}_{i}- \hat{\mathbf{x}}_M, \hat{\mathbf{p}}_{i})\frac{\hat{H}_{i}}{c} \right] + \Delta(\hat{\mathbf{x}}_1- \hat{\mathbf{x}}_M, \hat{\mathbf{p}}_1)\frac{\hat{H}_1}{c} + \hat{p}_0^M $ and the lowercase latin letters label all particles except for the particle serving as the QRF, i.e., $i= 2, \cdots, N$. The action of the operator $\hat{\mathcal{T}}_1$ on the constraints is
\begin{equation}
	\begin{split}
		\hat{\mathcal{T}}_1 \hat{f}^0 \hat{\mathcal{T}}_1^\dagger =& \sqrt{g_{00}(\hat{\mathbf{q}}_M)}\hat{k}^1_0; \qquad \hat{\mathcal{T}}_1 \hat{f}^1 \hat{\mathcal{T}}_1^\dagger = \hat{\mathbf{k}}_1; \qquad \hat{\mathcal{T}}_1 \hat{C}_i \hat{\mathcal{T}}_1^\dagger = \hat{C}'_i \,\,\, i\neq 1;\\
		 \hat{\mathcal{T}}_1 \hat{C}_1 \hat{\mathcal{T}}_1^\dagger = & \hat{k}_0^1 - \left\lbrace \sum_{i}\left[ \hat{k}_0^{i} + \Delta'(\hat{\mathbf{q}}_{i}, \hat{\mathbf{q}}_M, \hat{\mathbf{k}}_{i})\frac{\hat{H}_{i}}{c} \right] + \sqrt{g^{00}(\hat{\mathbf{q}}_M)}\hat{k}_0^M  \right\rbrace +\\
		 & - m_1 c\sqrt{1 + \frac{\left( \hat{\mathbf{k}}_1 - \sum_{i} \hat{\mathbf{k}}_{i} - \hat{\mathbf{k}}_M \right)^2}{m_1^2 c^2}}- \left(1 + \frac{\left( \hat{\mathbf{k}}_1 - \sum_{i} \hat{\mathbf{k}}_{i} - \hat{\mathbf{k}}_M \right)^2}{m_1^2 c^2}\right)^{-1/2} \frac{\hat{H}_1}{c},
	\end{split}
\end{equation}
where $\hat{C}'_i$ has been defined in the main text as $\hat{C}'_i = \sqrt{g'^{00}(\hat{\mathbf{q}}_i, \hat{\mathbf{q}}_M)} \hat{k}_0^i - \hat{\omega}_k^i$, $\Delta'(\hat{\mathbf{q}}_{i}, \hat{\mathbf{q}}_M, \hat{\mathbf{k}}_{i}) = \sqrt{g'_{00}(\hat{\mathbf{q}}_i, \hat{\mathbf{q}}_M)} \left( 1 + \frac{\hat{\mathbf{k}}_i^2}{m_i^2c^2} \right)^{-1/2}$, and the transformed metric field is
\begin{equation}
	g'_{00}(\hat{\mathbf{q}}_i, \hat{\mathbf{q}}_M) = \frac{g_{00} (\hat{\mathbf{q}}_i - \hat{\mathbf{q}}_M )}{g_{00} (\hat{\mathbf{q}}_M)}.
\end{equation}

The relational state from the perspective of system $1$ is then obtained as $\ket{\psi}^{(1)}=\bra{q_1 =0} \hat{\mathcal{T}}_1 \ket{\Psi}_{ph}$ 

By defining the internal time operator $\hat{T}_I$ such that $\left[ \hat{T}_I, \hat{H}_J \right]= i \hbar \delta_{IJ}$ and the state of the internal clock of each particle $\ket{\tau_I}$ as the one satisfying the relation $\hat{T}_I \ket{\tau_I} = \tau_I \ket{\tau_I}$, we find 
\begin{equation}
	\ket{\psi}^{(1)} = 2\pi\hbar \left(1 - \frac{\Phi(\mathbf{\hat{q}}_M)}{c^2}\right) \int d^N \mathcal{N}e^{\frac{\mi}{\hbar}\mathcal{N}_{i}\hat{C}'_{i}} e^{-\frac{\mi}{\hbar} \frac{\hat{\gamma}_{\Sigma k,1}^{-1}}{c}\left( \hat{K}^{(1)} +\hat{H}_1\right)\mathcal{N}_1} \bra{q_1=0} \hat{\Pi}_0 \mathcal{\hat{T}}_1\ket{\phi} ,
\end{equation}
where  $\hat{K}^{(1)}$ is the operator encoding the relational dynamics of particles $2, \cdots, N$ from the point of view of particle $1$, i.e.,
\begin{equation} \label{Appeq:K1}
	\hat{K}^{(1)} = \hat{\gamma}_{\Sigma k,1} \left[\sum_{i}c \hat{k}_0^{i} + c \sqrt{g^{00}(\hat{\mathbf{q}}_M)}\hat{k}_0^M \right] + \sum_{i} \sqrt{g'_{00}(\hat{\mathbf{q}}_{i}, \hat{\mathbf{q}}_M)} \frac{\hat{\gamma}_{i}^{-1}}{\hat{\gamma}_{\Sigma k,1}^{-1}}\hat{H}_{i} + m_1 c^2 \hat{\gamma}_{\Sigma k,1}^2,
\end{equation}
$\hat{\gamma}_{i} = \sqrt{1 + \frac{\hat{\mathbf{k}}_{i}^2}{m_{i}^2 c^2}}$, $\hat{\gamma}_{\Sigma k,1} = \sqrt{1 + \frac{\left(\sum_{i}\hat{\mathbf{k}}_{i} + \hat{\mathbf{k}}_M\right)^2}{m_{1}^2 c^2}}$, and $\hat{\Pi}_0 = \ket{k_1=0} \bra{k_1=0}$. We now define 
\begin{equation} \label{Appeq:phibar1}
	\ket{\phi_{/1}}= 2\pi\hbar \left(1 - \frac{\Phi(\mathbf{\hat{q}}_M)}{c^2}\right) \bra{q_1=0} \hat{\Pi}_0 \mathcal{\hat{T}}_1\ket{\phi}
\end{equation}
and act with the operator $\hat{H}_1$ on the internal state of clock $1$. We then rewrite 
\begin{equation}
	\ket{\psi}^{(1)} = \int d \tau_1 d^{N-1} \mathcal{N} e^{\frac{\mi}{\hbar}\mathcal{N}_{i}\hat{C}'_{i}} e^{-\frac{\mi}{\hbar} \hat{K}^{(1)} \tau_1} \ket{\phi^{(1)}_0} \ket{\tau_1},
\end{equation}
where 
\begin{equation}
	\begin{split}
		\ket{\phi^{(1)}_0} = &\int d t_1 \Pi_i [d^2 k_i dE_i]d^2 k_M e^{\frac{\mi}{\hbar} \hat{K}^{(1)} t_1} \phi_{/1}(k_2, \cdots, k_N, k_M, t_1, E_2, \cdots, E_N)\times\\
		& \times \ket{k_2, \cdots, k_N, k_M} \ket{E_2, \cdots, E_N}.
	\end{split}
\end{equation}

Following a similar procedure to the one outlined above, it is possible solve all the constraints and to cast the previous expression as a ``history state'' in the sense of Refs.~\cite{castro2020quantum, giacomini2020einstein} evolving unitarily as
\begin{equation}
	\ket{\psi}^{(1)} = \int d \tau_1 e^{-\frac{\mi}{\hbar} \hat{H}^{(1)} \tau_1} \ket{\psi^{(1)}_0} \ket{\tau_1},
\end{equation}
where the Hamiltonian $\hat{H}^{(1)}$ is
\begin{equation}
	\hat{H}^{(1)}= \hat{\gamma}_{\Sigma k,1} \sum_i \sqrt{g'_{00}(\hat{\mathbf{q}}_i, \hat{\mathbf{q}}_M)}\left\lbrace c\hat{\omega}_k^i + \hat{\gamma}_i^{-1} \hat{H}_i \right\rbrace + c \hat{\gamma}_{\Sigma k,1} \sqrt{g^{00}(\hat{\mathbf{q}}_M)} \hat{k}_0^M + m_1 c^2 \hat{\gamma}_{\Sigma k,1}^2,
\end{equation}
and 
\begin{equation}
	\begin{split}
		\ket{\psi^{(1)}_0} = &\int \Pi_i \left(d\mu(q_i)d q'^{0}_i dE_i \sqrt{g'_{00}(\mathbf{q}_i, \mathbf{q}_M)}\right)d^2 q_M e^{\frac{\mi}{\hbar} \sum_i (q'^0_i - q^0_i)\sqrt{g'_{00}(\mathbf{q}_i, \mathbf{q}_M)} \hat{\omega}_k^i} \times \\
		& \times \phi^{(1)}_0\left( q_2, \cdots, q_N, q_M, E_2, \cdots, E_N\right) \ket{q'^0_2, \mathbf{q}_2, \cdots, q'^0_N, \mathbf{q}_N}\ket{q_M} \ket{E_2 \cdots E_N},
	\end{split}
\end{equation}
which reproduces the result of the main text.

The generalisation of this technique to general gravitational fields requires dealing with second-class constraints, and is thus beyond the scope of this work.

\section{Limiting cases of the general $N$-particle model}
\label{App:Limits}

In the following, we are going to study some relevant limits of the model introduced in the main text. In particular, we will focus on a set of quantum free particles moving slowly compared to the speed of light (Galilean case), on a set of quantum special-relativistic particles, and on a set of quantum Galilean particles in an external weak gravitational field (Newtonian case).

We remind the reader that operators and vectors with no indices are, unless differently specified, operators and two-vectors in spacetime, while the spatial component of operators and vectors is boldface. Greek letters label spacetime indices, capital Latin letters label all the particles, e.g., $I = 1, \cdots, N$, and lowercase Latin letters label all the particles except the one serving as the QRF, e.g., $i=2, \cdots, N$.

\subsection{Galilean case}
\label{SubApp:LimitsGalilean}

When the particles move slowly compared to the speed of light and are free, the contraints introduced in the main text reduce to
\begin{align}
	& \hat{C}_I = \hat{p}_I^0 - \frac{\hat{\mathbf{p}}^2_I}{2m_Ic},\\
	& \hat{f}^0 = \sum_{I=1}^N \left(\hat{p}_I^0 + \frac{\hat{H}_I}{c}\right),\\
	& \hat{f}^1 = \sum_{I=1}^N \hat{\mathbf{p}}_I,
\end{align}
 where here we have neglected particle $M$ because it plays no role. In addition, we eliminated the constant term $m_I c$ in the expansion of $\hat{C}_I$, because it simply amounts to a rescaling of the zero component of the momentum. 
 
We write the state in the Physical Hilbert space as
\begin{equation}
	\ket{\Psi}_{ph} \propto \int d^N \mathcal{N} d^2 z e^{\frac{\mi}{\hbar}\mathcal{N}_i \hat{C}_i}e^{\frac{\mi}{\hbar}z_\mu \hat{f}^\mu}\ket{\phi},
\end{equation}
where $\ket{\phi}$ is expanded in momentum basis as
\begin{equation}
	\ket{\phi} = \int \Pi_{I} \left[d^2 p_I d E_I\right] \phi(p_1, \cdots, p_N, E_1, \cdots, E_N)\ket{p_1, \cdots, p_N}\ket{E_1, \cdots, E_N}.
\end{equation}
 
In this case, the operator $\hat{\mathcal{T}}_1$ simplifies to
\begin{equation}
	\hat{\mathcal{T}}_1 = e^{\frac{\mi}{\hbar}\hat{\mathbf{x}}_1 \sum_{i} \hat{\mathbf{p}}_{i}} e^{\frac{\mi}{\hbar}\hat{x}_1^0 \left[ \sum_{i}\left(\hat{p}_0^{i} + \frac{\hat{H}_{i}}{c} \right) + \frac{\hat{H}_1}{c}\right]}, 
\end{equation}
and the constraints are mapped to
\begin{equation}
	\begin{split}
		&\hat{\mathcal{T}}_1 \hat{f}^0 \hat{\mathcal{T}}_1^\dagger = \hat{k}^1_0; \qquad \hat{\mathcal{T}}_1 \hat{f}^1 \hat{\mathcal{T}}_1^\dagger = \hat{\mathbf{k}}_1; \qquad \hat{\mathcal{T}}_1 \hat{C}_i \hat{\mathcal{T}}_1^\dagger = \hat{C}_i \,\,\, i\neq 1;\\
		& \hat{\mathcal{T}}_1 \hat{C}_1 \hat{\mathcal{T}}_1^\dagger = \hat{k}_0^1 - \sum_{i}\left(\hat{k}_0^{i} + \frac{\hat{H}_{i}}{c} \right) - \frac{\hat{H}_1}{c} - \frac{\left( \hat{\mathbf{k}}_1 - \sum_{i} \hat{\mathbf{k}}_{i}\right)^2}{2m_1}.
	\end{split}
\end{equation}

Defining the relational state from the perspective of particle $1$ to be $\ket{\psi^{(1)}} = {}_1\bra{q_1 = 0} \mathcal{T}_1 \ket{\Psi}_{ph}$, we find
\begin{equation}
	\ket{\psi}^{(1)} = \int d\tau_1 d^{N-1}\mathcal{N} \ket{\tau_1} e^{\frac{\mi}{\hbar}\mathcal{N}_{i} \hat{C}_{i}} e^{-\frac{\mi}{\hbar}\hat{K}^{(1)}\tau_1}\ket{\phi^{(1)}_0} ,
\end{equation}
where 
\begin{equation}
	\hat{K}^{(1)} = \sum_{i} \left(\hat{k}_0^{i} + \frac{\hat{H}_{i}}{c} \right)+ \frac{\left( \sum_{i} \hat{\mathbf{k}}_{i}\right)^2}{2m_1},
\end{equation}
and 
\begin{equation}
	\begin{split}
		&\ket{\phi^{(1)}_0} =  \frac{1}{\sqrt{2\pi\hbar}}\int dt_1 dE_1 \Pi_{i}(d^2 k_{i} dE_{i}) e^{\frac{\mi}{\hbar}(\hat{K}^{(1)} + E_1) t_1}\phi\left(\bar{k}_1, k_2, \cdots, k_N, E_1, \cdots, E_N\right)\times \\
	& \hspace{2cm}\times\ket{k_2 \cdots k_N} \ket{E_2 \cdots E_N}.
	\end{split}
\end{equation}
By solving the constraint we have rewritten the two-vector $\bar{k}_1 = \left(- \sum_{i}\left(k_0^{i} + \frac{E_{i}}{c} \right)-\frac{E_1}{c}, -\sum_{i} \mathbf{k}_{i}\right)$. Notice that we can easily recover the framework of standard quantum mechanics by integrating over the $N_{i}$ variables. After this operation, the state $\ket{\psi^{(1)}}$ can be rewritten as the ``history state''
\begin{equation}
	\ket{\psi}^{(1)}  = \int d\tau_1 e^{-\frac{\mi}{\hbar}\hat{H}^{(1)}\tau_1}\ket{\psi^{(1)}_0} \ket{\tau_1},
\end{equation}
where
\begin{align*}
	 &\hat{H}^{(1)} = \sum_{i} \frac{\hat{\mathbf{k}}_{i}^2}{2m_{i}} + \frac{(\sum_{{i}}\hat{\mathbf{k}}_{i})^2}{2m_1} + \sum_{i} \hat{H}_{i},\\
	\begin{split}
		&\ket{\psi^{(1)}_0} = \int \Pi_{i}(d\mathbf{k}_{i} dE_{i}) \phi^{(1)}_0\left(\frac{\mathbf{k}_2^2}{2m_2}, \mathbf{k}_2, \cdots, \frac{\mathbf{k}_N^2}{2m_N}, \mathbf{k}_N, E_1, \cdots, E_N\right)\times \\
	&\hspace{2cm} \times \ket{\frac{\mathbf{k}_2^2}{2m_2}, \mathbf{k}_2, \cdots, \frac{\mathbf{k}_N^2}{2m_N}, \mathbf{k}_N} \ket{E_2 \cdots E_N}.
	\end{split}
\end{align*}

\subsection{Special-relativistic case}
\label{SubApp:LimitsSpecRel}

In this section, we consider a set of N special-relativistic particles moving freely. The set of constraints is
\begin{align}
	& \hat{C}_I =  \hat{p}_I^0 -\hat{\omega}_p^I,\\
	& \hat{f}^0 = \sum_{I=1}^N \left(\hat{p}_I^0 +  \hat{\gamma}_I^{-1}\frac{\hat{H}_I}{c}\right),\\
	& \hat{f}^1 = \sum_{I=1}^N \hat{\mathbf{p}}_I,
\end{align}
 where we have defined $\hat{\omega}_p^I = m_I c^2 \sqrt{1+ \frac{\hat{\mathbf{p}}_I^2}{m_I^2 c^2}}$ and $\hat{\gamma}_I = \sqrt{1+ \frac{\hat{\mathbf{p}}_I^2}{m_I^2 c^2}}$. We write the state in the Physical Hilbert space as
\begin{equation}
	\ket{\Psi}_{ph} \propto \int d^N \mathcal{N} d^2 z e^{\frac{\mi}{\hbar}\mathcal{N}_i \hat{C}_i}e^{\frac{\mi}{\hbar}z_\mu \hat{f}^\mu}\ket{\phi},
\end{equation}
where $\ket{\phi}$ is expanded in momentum basis as
\begin{equation}
	\ket{\phi} = \int \Pi_{I} \left[d^2 p_I d E_I\right] \phi(p_1, \cdots, p_N, E_1, \cdots, E_N)\ket{p_1, \cdots, p_N}\ket{E_1, \cdots, E_N},
\end{equation}
 
In this case, the trivialisation operator is
\begin{equation}
	\hat{\mathcal{T}}_1 = e^{\frac{\mi}{\hbar}\hat{\mathbf{x}}_1 \sum_{i} \hat{\mathbf{p}}_{i}} e^{\frac{\mi}{\hbar}\hat{x}_1^0 \left[ \sum_{i}\left(\hat{p}_0^{i} + \hat{\gamma}_i^{-1}\frac{\hat{H}_{i}}{c} \right) +\, \hat{\gamma}_1^{-1} \frac{\hat{H}_1}{c}\right]}, 
\end{equation}
and the constraints are mapped to
\begin{equation}
	\begin{split}
		&\hat{\mathcal{T}}_1 \hat{f}^0 \hat{\mathcal{T}}_1^\dagger = \hat{k}^1_0; \qquad \hat{\mathcal{T}}_1 \hat{f}^1 \hat{\mathcal{T}}_1^\dagger = \hat{\mathbf{k}}_1; \qquad \hat{\mathcal{T}}_1 \hat{C}_i \hat{\mathcal{T}}_1^\dagger = \hat{C}_i \,\,\, i\neq 1;\\
		& \hat{\mathcal{T}}_1 \hat{C}_1 \hat{\mathcal{T}}_1^\dagger = \hat{k}_0^1 - \sum_{i}\left(\hat{k}_0^{i} + \hat{\gamma}_i^{-1}\frac{\hat{H}_{i}}{c} \right) - \left(1 + \frac{(\hat{\mathbf{k}}_1 - \sum_i \hat{\mathbf{k}}_i)^2}{m_1^2 c^2} \right)^{-1/2} \frac{\hat{H}_1}{c} - m_1 c \sqrt{1 + \frac{(\hat{\mathbf{k}}_1 - \sum_i \hat{\mathbf{k}}_i)^2}{m_1^2 c^2}}.
	\end{split}
\end{equation}

Following analogous steps to the Galilean case, we define
$\ket{\psi}^{(1)} = \bra{q_1 = 0} \hat{\mathcal{T}}_1 \ket{\Psi}_{ph}$, where
\begin{equation}
	\ket{\psi}^{(1)} = \int d\tau_1 d^{N-1}\mathcal{N} \ket{\tau_1} e^{\frac{\mi}{\hbar}\mathcal{N}_{i} \hat{C}_{i}} e^{-\frac{\mi}{\hbar}\hat{K}^{(1)}\tau_1}\ket{\phi^{(1)}_0},
\end{equation}
where 
\begin{equation}
	\hat{K}^{(1)} = \hat{\gamma}_{\Sigma k, 1} \left[ \sum_{i} \left(c\hat{k}_0^{i} + \hat{\gamma}_i^{-1}\hat{H}_{i} \right)+ c \hat{\omega}_{\Sigma k}^1 \right],
\end{equation}
and
\begin{equation}
	\begin{split}
		&\ket{\phi^{(1)}_0} =  \frac{1}{\sqrt{2\pi\hbar}}\int \Pi_{i}(d^2 k_{i} dE_{i})dt_1 dE_1  \gamma_{\Sigma k,1} e^{\frac{\mi}{\hbar}(\hat{K}^{(1)} + E_1) t_1}\phi\left(\bar{k}_1, k_2, \cdots, k_N, E_1, \cdots, E_N\right)\times \\
	& \hspace{2cm}\times\ket{k_2 \cdots k_N} \ket{E_2 \cdots E_N}.
	\end{split}
\end{equation}
Here, we have defined $\hat{\gamma}_{\Sigma k,1} = \sqrt{1 + \frac{(\sum_i \hat{\mathbf{k}}_i)^2}{m_1^2 c^2}}$ and $\hat{\omega}_{\Sigma k}^1 = m_1 c \sqrt{1 + \frac{( \sum_i \hat{\mathbf{k}}_i)^2}{m_1^2 c^2}}$ and the two-vector $\bar{k}_1 = \left(- \sum_{i}\left(k_0^{i} + \gamma_i^{-1}\frac{E_{i}}{c} \right)-\gamma_{\Sigma k, 1}^{-1}\frac{E_1}{c}, -\sum_{i} \mathbf{k}_{i}\right)$.

It is again possible, with a similar procedure to the Galilean case, to recover the dynamical evolution of a quantum relativistic particle. We find
\begin{equation}
	\ket{\psi}^{(1)}  = \int d\tau_1 e^{-\frac{\mi}{\hbar}\hat{H}^{(1)}\tau_1}\ket{\psi^{(1)}_0} \ket{\tau_1},
\end{equation}
where
\begin{align*}
	 &\hat{H}^{(1)} = \hat{\gamma}_{\Sigma k, 1} \left[ \sum_{i} \left(c\hat{\omega}_k^{i} + \hat{\gamma}_i^{-1}\hat{H}_{i} \right)+ c \hat{\omega}_{\Sigma k}^1 \right],\\
	\begin{split}
		&\ket{\psi^{(1)}_0} = \int \Pi_{i}(d\mathbf{k}_{i} dE_{i}) \phi^{(1)}_0\left(\omega_k^2, \mathbf{k}_2, \cdots, \omega_k^N, \mathbf{k}_N , E_2, \cdots, E_N\right)\times \\
		& \hspace{2cm}\times \ket{\omega_k^2, \mathbf{k}_2, \cdots, \omega_k^N, \mathbf{k}_N} \ket{E_2 \cdots E_N}.
	\end{split}
\end{align*}

Note that the factor $\sqrt{1+ \frac{\left( \sum_{i\neq 1} \hat{k}_i\right)^{2}}{m_1^2 c^2}}$ multiplying the hamiltonian of the $N-1$ particles from the point of view of particle $1$ is the special-relativistic time dilation due to the fact that the QRF is moving in a superposition of special relativistic velocities. 

\subsection{Newtonian case}
\label{SubApp:LimitsNewtonian}

Finally, we consider the case in which a set of $N$ particles move in a Newtonian gravitational field produced by a mass $M$. The gravitational field in the Newtonian limit is the same as in the main text
\begin{equation}
	\begin{split}
		& g_{00} = 1 + \frac{2\Phi(\mathbf{x}- \mathbf{x}_M)}{c^2};\\
		& g_{01} = g_{10} = 0;\\
		& g_{11} = -1,
	\end{split}
\end{equation}
where $\Phi(\mathbf{x})$ is the Newtonian potential due to the mass $m_M$ of the system $M$ and $|\Phi(\mathbf{x})|/c^2 \ll 1$ in the spacetime region considered.

The set of constraints for this situation are 
\begin{equation}
	\begin{split}
		& \hat{C}_I = \sqrt{g^{00}(\hat{\mathbf{x}}_I - \hat{\mathbf{x}}_M)}\hat{p}^I_0 -m_I c- \frac{\hat{\mathbf{p}}^2_I}{2m_I c} \qquad \text{for}\qquad I=1, \cdots, N;\\
		& \hat{f}^0 = \sum_{I=1}^N \left[\hat{p}^I_0 +  \sqrt{g_{00}(\hat{\mathbf{x}}_I - \hat{\mathbf{x}}_M)} \frac{\hat{H}_I}{c}\right] + \hat{p}_0^M;\\
		& \hat{f}^1 = \sum_{I=1}^N \hat{\mathbf{p}}_I + \hat{\mathbf{p}}_M.
	\end{split}
\end{equation}

Notice that, since this is a special case of the case considered in the main text, the constraints are still first-class to our order of approximation.

The state in the Physical Hilbert space is
\begin{equation}
	\ket{\Psi}_{ph} \propto \int d^N \mathcal{N} d^2 z e^{\frac{\mi}{\hbar}\mathcal{N}_i \hat{C}_i}e^{\frac{\mi}{\hbar}z_\mu \hat{f}^\mu}\ket{\phi},
\end{equation}
where, in this case, it is convenient to expand $\ket{\phi}$ in position basis as
\begin{equation}
	\ket{\phi} = \int \Pi_{I} \left[d \mu(x_I) d E_I\right]d^2 x_M \phi(x_1, \cdots, x_N, x_M, E_1, \cdots, E_N)\ket{x_1, \cdots, x_N, x_M}\ket{E_1, \cdots, E_N},
\end{equation}
where $d \mu(x_I) = \sqrt{g_{00}(\mathbf{x}_I-\mathbf{x}_M)} d^2 x_I$ is the covariant integration measure. 

We define the operator $\hat{\mathcal{T}}_1$ as
\begin{equation}
	\hat{\mathcal{T}}_1 = e^{\frac{i}{\hbar}\frac{\log\sqrt{g_{00}(\hat{\mathbf{x}}_M)}}{2}\sum_{I=1}^N(\hat{x}^0_I \hat{p}_0^I + \hat{p}_0^I  \hat{x}^0_I)} e^{\frac{\mi}{\hbar}\hat{\mathbf{x}}_1 \left(\hat{f}^1 - \hat{\mathbf{p}}^1 \right)} e^{\frac{\mi}{\hbar}\hat{x}_1^0 \left( \hat{f}^0 - \hat{p}^1_0 \right)}, 
\end{equation}
where in this case $\hat{f}^1 - \hat{\mathbf{p}}^1 = \sum_{i} \hat{\mathbf{p}}_{i} +\hat{\mathbf{p}}_M $ and $\hat{f}^0 - \hat{p}^1_0 =  \sum_{i}\left[ \hat{p}_0^{i} + \sqrt{g_{00}(\hat{\mathbf{x}}_{i}- \hat{\mathbf{x}}_M)}\frac{\hat{H}_{i}}{c} \right] + \sqrt{g_{00}(\hat{\mathbf{x}}_{1}- \hat{\mathbf{x}}_M)}\frac{\hat{H}_1}{c} + \hat{p}_0^M $. 

The action of the operator $\hat{\mathcal{T}}_1$ on the constraints is
\begin{equation}
	\begin{split}
		&\hat{\mathcal{T}}_1 \hat{f}^0 \hat{\mathcal{T}}_1^\dagger = \sqrt{g_{00}(\hat{\mathbf{q}}_M)} \hat{k}^1_0; \qquad \hat{\mathcal{T}}_1 \hat{f}^1 \hat{\mathcal{T}}_1^\dagger = \hat{\mathbf{k}}^1; \qquad \hat{\mathcal{T}}_1 \hat{C}_i \hat{\mathcal{T}}_1^\dagger = \hat{C}'_i \,\,\, i\neq 1;\\
		& \hat{\mathcal{T}}_1 \hat{C}_1 \hat{\mathcal{T}}_1^\dagger =  \hat{k}_0^1 - \left[\sum_{i}\left( \hat{k}_0^{i} + \sqrt{g'_{00}(\hat{\mathbf{q}}_{i}, \hat{\mathbf{q}}_M)}\frac{\hat{H}_{i}}{c} \right) + \sqrt{g^{00}(\hat{\mathbf{q}}_M)} \hat{k}_0^M  \right] +\\
		&\hspace{2cm} -m_1 c- \frac{(\hat{\mathbf{k}}_1 -\sum_i \hat{\mathbf{k}}_i - \hat{\mathbf{k}}_M)^2}{2m_1c} - \frac{\hat{H}_1}{c},
	\end{split}
\end{equation}
where $\hat{C}'_i = \sqrt{g'_{00}(\hat{\mathbf{q}}_i, \hat{\mathbf{q}}_M)}\hat{k}^i_0 -m_i c- \frac{\hat{\mathbf{k}}^2_i}{2m_i c}$ and $g'^{00}(\hat{\mathbf{q}}_i, \hat{\mathbf{q}}_M) = \frac{g_{00} (\hat{\mathbf{q}}_i - \hat{\mathbf{q}}_M )}{g_{00} (\hat{\mathbf{q}}_M)}$ as in the main text.

Similarly to the other cases, the relational state from the perspective of system $1$ is $\ket{\psi}^{(1)}=\bra{q_1 =0} \hat{\mathcal{T}}_1 \ket{\Psi}_{ph}$. In order to find its explicit expression, we here define
\begin{equation}
	\begin{split}
		\ket{\tilde{\phi}} = \hat{\mathcal{T}}_1 \ket{\phi} = &\int d^2 k_1 \Pi_i\left[d\mu(q_i)dE_i \right]dE_1 d^2 q_M \tilde{\phi}(k_1, q_2, \cdots,q_N, q_M, t_1, E_2, \cdots, E_N)\times\\
		&\times \ket{k_1, q_2, \cdots,q_N, q_M} \ket{t_1} \ket{E_2, \cdots, E_N}.
	\end{split}
\end{equation}
We then find
\begin{equation}
	\ket{\psi}^{(1)} = \int d \tau_1 d^{N-1} \mathcal{N} e^{\frac{\mi}{\hbar}\mathcal{N}_{i}\hat{C}'_{i}} e^{-\frac{\mi}{\hbar} \hat{K}^{(1)} \tau_1} \ket{\phi^{(1)}_0} \ket{\tau_1},
\end{equation}
where 
\begin{equation}
	\begin{split}
		\ket{\phi^{(1)}_0}& =c\int dt_1 \Pi_i [d\mu(q_i) dE_i]d^2 q_M \sqrt{g^{00}(\mathbf{q}_M)} e^{\frac{\mi}{\hbar} \hat{K}^{(1)} t_1} \times \\
		& \times  \tilde{\phi}(k_1=0, q_2, \cdots, q_N, q_M, t_1, E_2 \cdots, E_N) \ket{q_2, \cdots, q_N, q_M} \ket{E_2, \cdots, E_N},
	\end{split}
\end{equation}
and $\hat{K}^{(1)}$ is the operator encoding the relational dynamics of particles $2, \cdots, N$ from the point of view of particle $1$, i.e.,
\begin{equation}
	\hat{K}^{(1)} =  \sum_{i}c \hat{k}_0^{i} + c \sqrt{g^{00}(\hat{\mathbf{q}}_M)} \hat{k}_0^M  + \sum_{i} \sqrt{g'_{00}(\hat{\mathbf{q}}_{i}, \hat{\mathbf{q}}_M)} \hat{H}_{i} + m_1 c^2 + \frac{(\sum_i \hat{\mathbf{k}}_i + \hat{\mathbf{k}}_M)^2}{2m_1}.
\end{equation}

With a similar procedure to the previous cases, we find that we can recover the usual dynamical evolution as
\begin{equation}
	\ket{\psi}^{(1)}  = \int d\tau_1 \ket{\tau_1} e^{-\frac{\mi}{\hbar}\hat{H}^{(1)}\tau_1}\ket{\psi^{(1)}_0} ,
\end{equation}
where the relational Hamiltonian in the perspective of particle $1$ is
\begin{equation}
	\hat{H}^{(1)} = \sqrt{g^{00}(\hat{\mathbf{q}}_M)} c \hat{k}_0^M + \sum_{i} \sqrt{g'_{00}(\hat{\mathbf{q}}_{i}, \hat{\mathbf{q}}_M)}\left[m_i c^2 + \frac{\hat{\mathbf{k}}_i^2}{2m_i} + \hat{H}_{i}\right] + m_1 c^2 + \frac{(\sum_i \hat{\mathbf{k}}_i + \hat{\mathbf{k}}_M)^2}{2m_1},
\end{equation}
and 
\begin{equation}
	\begin{split}
		&\ket{\psi^{(1)}_0} = \int \Pi_{i}\left[ d\mu(q_i) d q'^{0}_i \sqrt{g'_{00}(\mathbf{q}_i - \mathbf{q}_M)} dE_{i}\right] d^2 q_M e^{\frac{\mi}{\hbar} \sum_i (q'^0_i - q^0_i)\sqrt{g'_{00}(\mathbf{q}_i - \mathbf{q}_M)} \left[m_i c + \frac{\hat{\mathbf{k}}_i^2}{2m_i c} \right]} \times \\
		& \times \phi^{(1)}_0\left( q_2, \cdots, q_N, q_M, E_2, \cdots, E_N\right) \ket{q'^0_2, \mathbf{q}_2, \cdots, q'^0_N, \mathbf{q}_N}\ket{q_M} \ket{E_2 \cdots E_N}.
	\end{split}
\end{equation}

Notice that, if we expand the Hamiltonian to lowest order in the general relativistic corrections for the external degrees of freedom, while still allowing the internal degrees of freedom to have relativistic corrections, we obtain
\begin{equation}
	\begin{split}
		\hat{H}^{(1)} =& \sum_{i} \left\lbrace m_i c^2 + \frac{\hat{\mathbf{k}}_i^2}{2m_i} +m_i [\Phi(\hat{\mathbf{q}}_i- \hat{\mathbf{q}}_M) - \Phi(\hat{\mathbf{q}}_M)] +\left[1+ \frac{[\Phi(\hat{\mathbf{q}}_i- \hat{\mathbf{q}}_M) - \Phi(\hat{\mathbf{q}}_M)]}{c^2}\right]\hat{H}_{i}\right\rbrace +\\
		&+ m_1 c^2 + \frac{(\sum_i \hat{\mathbf{k}}_i + \hat{\mathbf{k}}_M)^2}{2m_1} + \sqrt{g^{00}(\hat{\mathbf{q}}_M)} c \hat{k}_0^M. 
	\end{split}
\end{equation}
As a result, we see that the potential in the QRF of particle $1$ is the difference between the gravitational potential between particle $i$ and the mass $M$ and particle $1$ and the mass $M$. This is the straightforward generalisation of the standard reference frame description. 

\section{Explicit calculation of the ``history state'' with a quantum measurement}
\label{App:CalcMeas}

The calculation of the history state in the case with a measurement parallels the one outlined in Appendix~\ref{App:calculation}. With a fully analogous procedure, fixing $N=2$, and choosing the constraints of Eq.~\eqref{eq:CconstraintMeas}
\begin{equation}
	\begin{split}
		& \hat{C}_I = \sqrt{g^{00}(\hat{\mathbf{x}}_I - \hat{\mathbf{x}}_M)}\hat{p}^I_0 -\hat{\omega}_p^I \qquad \text{for}\qquad I=1, 2;\\
		& \hat{f}_Q^0 = \hat{p}^1_0 + \hat{p}^2_0 + \hat{p}_0^M + \Delta(\hat{\mathbf{x}}_1 - \hat{\mathbf{x}}_M, \hat{\mathbf{p}}_1) \frac{\hat{H}_1}{c} + \Delta(\hat{\mathbf{x}}_2 - \hat{\mathbf{x}}_M, \hat{\mathbf{p}}_2)\left[\frac{\hat{H}_2}{c}+\delta(\hat{T}_2 - \tau_2*) \frac{\hat{Q}_2}{c}\right];\\
		& \hat{f}^1 = \hat{\mathbf{p}}_1 + \hat{\mathbf{p}}_2 + \hat{\mathbf{p}}_M,
	\end{split}
\end{equation}
where $\hat{Q}_2$ is an observable commuting with every constraint, and the operator 
\begin{equation}
	\hat{\mathcal{T}}_{Q,1} = e^{\frac{i}{\hbar}\frac{\log\sqrt{g_{00}(\hat{\mathbf{x}}_M)}}{2}\sum_{I=1}^N(\hat{x}^0_I \hat{p}_0^I + \hat{p}_0^I  \hat{x}^0_I)} e^{\frac{\mi}{\hbar}\hat{\mathbf{x}}_1 \left(\hat{f}^1 - \hat{\mathbf{p}}^1 \right)} e^{\frac{\mi}{\hbar}\hat{x}_1^0 \left( \hat{f}_Q^0 - \hat{p}^1_0 \right)},
\end{equation}
we find that the history state can be cast in the form
\begin{equation} \label{Appeq:psi1Meas}
	\ket{\psi}^{(1)} = \int d \tau_1 d^{N-1} \mathcal{N} e^{\frac{\mi}{\hbar}\mathcal{N}_{i}\hat{C}'_{i}} e^{-\frac{\mi}{\hbar} \hat{K}_Q^{(1)} \tau_1} \ket{\phi^{(1)}_0} \ket{\tau_1},
\end{equation}
where $\hat{C}'_{i}$ is the same as in the main text and in Appendix~\ref{App:calculation}. In this case, we have that
\begin{equation}
	\hat{K}_Q^{(1)} = \hat{K}^{(1)} + \sqrt{g'_{00}(\hat{\mathbf{q}}_{2}, \hat{\mathbf{q}}_M)}\frac{\hat{\gamma}_{2}^{-1}}{\hat{\gamma}_{\Sigma k,1}^{-1}}\delta(\hat{T}_2 - \tau_2^*)\hat{Q}_2,
\end{equation}
where $\hat{K}^{(1)}$ was defined in Eq.~\eqref{Appeq:K1} for a general number of particles $N$ (which here we have fixed to $N=2$). We also recall that we have defined $\hat{\gamma}_{2} = \sqrt{1 + \frac{\hat{\mathbf{k}}_{2}^2}{m_{2}^2 c^2}}$ and $\hat{\gamma}_{\Sigma k,1} = \sqrt{1 + \frac{\left(\hat{\mathbf{k}}_{2} + \hat{\mathbf{k}}_M\right)^2}{m_{1}^2 c^2}}$. Finally, we also write
\begin{equation}
	\ket{\phi^{(1)}_0} = c \int d t_1 d^2 k_2 dt_2 d^2 k_M \gamma_{\Sigma k,1} e^{\frac{\mi}{\hbar} \hat{K}_Q^{(1)} t_1} \phi_{/1}(k_2, k_M, t_1, t_2)\ket{k_2, k_M, t_2}
\end{equation}
where the state $\ket{\phi_{/1}}$ has the same expression as in Eq.~\eqref{Appeq:phibar1} with the operator $\hat{\mathcal{T}}_{Q,1}$ instead of $\hat{\mathcal{T}}_1$. We now consider the following equality, proved in Ref.~\cite{castro2020quantum}, 
\begin{equation}
	e^{-\frac{\mi}{\hbar}\alpha[ \hat{H}_I + \hat{f}(\hat{T}_I)]} = e^{-\frac{\mi}{\hbar}\alpha \hat{H}_I} \overleftarrow{T}\left\lbrace e^{-\frac{\mi}{\hbar}\int_0^1 ds \alpha \hat{f}(\hat{T}_I + \alpha s)} \right\rbrace,
\end{equation}
where $[\hat{T}_I, \hat{H}_I] = \mi \hbar$ and $\hat{f}(\hat{T}_I)$ is an operator-valued function, and we slightly generalise it to
\begin{equation}
	e^{-\frac{\mi}{\hbar}\alpha[\hat{K}_R + \hat{H}_I + \hat{f}(\hat{T}_I)]} = e^{-\frac{\mi}{\hbar}\alpha \hat{H}_I} \overleftarrow{T}\left\lbrace e^{-\frac{\mi}{\hbar}\int_0^1 ds \alpha \left[\hat{K}_R +\hat{f}(\hat{T}_I + \alpha s)\right]} \right\rbrace,
\end{equation}
where $\hat{K}_R$ is an arbitrary hermitian operator which does not necessarily commute with $\hat{f}(\hat{T}_I)$. Then, the expression of Eq:~\eqref{Appeq:psi1Meas} can be cast as 
\begin{equation}
	\ket{\psi}^{(1)} = \int d \tau_1 \overleftarrow{T}\left\lbrace  e^{-\frac{\mi}{\hbar}\int_0^{\tau_1} ds \left[\hat{H}^{(1)}+ \Delta_{12} \delta(\hat{T}_2 + \Delta_{12}s - \tau_2^*)\hat{Q}_2 \right]} \right\rbrace \ket{\psi^{(1)}_0} \ket{\tau_1},
\end{equation}
where $\Delta_{12} = \sqrt{g'_{00}(\hat{\mathbf{q}}_{2}, \hat{\mathbf{q}}_M)}\frac{\hat{\gamma}_{2}^{-1}}{\hat{\gamma}_{\Sigma k,1}^{-1}}$. The Hamiltonian $\hat{H}^{(1)}$ is the same as in the main text
\begin{equation}
	\hat{H}^{(1)}= \hat{\gamma}_{\Sigma k,1} \sum_i \sqrt{g'_{00}(\hat{\mathbf{q}}_i, \hat{\mathbf{q}}_M)}\left\lbrace c\hat{\omega}_k^i + \hat{\gamma}_i^{-1} \hat{H}_i \right\rbrace + c \hat{\gamma}_{\Sigma k,1} \sqrt{g^{00}(\hat{\mathbf{q}}_M)} \hat{k}_0^M + m_1 c^2 \hat{\gamma}_{\Sigma k,1}^2,
\end{equation}
and 
\begin{equation}
	\begin{split}
		\ket{\psi^{(1)}_0} = &\int d\mu(q_2)d q'^{0}_2 dt_2 d^2 q_M \sqrt{g'_{00}(\mathbf{q}_2 - \mathbf{q}_M)}  e^{\frac{\mi}{\hbar} (q'^0_2 - q^0_2)\sqrt{g'_{00}(\mathbf{q}_2 - \mathbf{q}_M)} \hat{\omega}_k^2} \times \\
		& \times \phi^{(1)}_0\left( q_2, q_M, t_2 \right) \ket{q'^0_2, \mathbf{q}_2}\ket{q_M} \ket{t_2},
	\end{split}
\end{equation}

\bibliography{biblioCovQRF}{}

\end{document}